\documentclass[structabstract]{aa}  
\usepackage[]{natbib}
\usepackage{lscape}
\usepackage{longtable}
\usepackage{threeparttable}
\usepackage{graphicx}
\usepackage{amsmath}
\usepackage{txfonts}
\renewcommand{\arraystretch}{1.4}
\begin{document}

   \title{A star-forming dwarf galaxy candidate in the halo of NGC 4634}

   \author{Y. Stein,
                                        \inst{1,2}
           D.J. Bomans, 
                 \inst{2}
           P. Kamphuis,
                \inst{2,3}
           E. J\"utte,
                \inst{2}
           M. Langener,
           \inst{2}
           \and
           R.-J. Dettmar \inst{2}
           }

   \institute{Universit\'e de Strasbourg, CNRS, Observatoire astronomique de Strasbourg, UMR 7550, F-67000 Strasbourg, France; \email{stein@astro.rub.de}
                                                \and
                                    Astronomisches Institut (AIRUB), Ruhr-Universit\"at Bochum, Universit\"atsstrasse 150, 44801 Bochum, Germany
              \and
              National Centre for Radio Astrophysics, TIFR, Ganeshkhind, Pune 411007, India
               }

   \date{Accepted July 25, 2018}

  \abstract
   {The halos of disk galaxies form a crucial connection between the galaxy disk and the intergalactic medium. Massive stars, H{\sc ii} regions, or dwarf galaxies located in the halos of galaxies are potential tracers of recent accretion and/or outflows of gas, and are additional contributors to the photon field and the gas phase metallicity.} 
   {We investigate the nature and origin of a star-forming dwarf galaxy candidate located in the halo of the edge-on Virgo galaxy NGC~4634 with a projected distance of 1.4 kpc and a H$\alpha$ star formation rate of $\sim 4.7 \times 10^{-3} \text{M}_\odot \text{yr}^{-1}$ in order to increase our understanding of these disk-halo processes.}  
   {With optical long-slit spectra we measured fluxes of optical nebula emission lines to derive the oxygen abundance 12~+~log(O/H) of an H{\sc ii} region in the disk of NGC~4634 and in the star-forming dwarf galaxy candidate. Abundances derived from optical long-slit data and from Hubble Space Telescope (HST) r-band data, H$\alpha$ data, Giant Metrewave Radio Telescope (GMRT) H{\sc i} data, and photometry of SDSS and GALEX data were used for further analysis. With additional probes of the luminosity--metallicity relation in the $B$-band from the H$\alpha$-luminosity, the H{\sc i} map, and the relative velocities, we are able to constrain a possible origin of the dwarf galaxy candidate.}
   {The high oxygen abundance (12 + log(O/H) $\approx$ 8.72) of the dwarf galaxy candidate leads to the conclusion that it was formed from pre-enriched material. Analysis of auxiliary data shows that the dwarf galaxy candidate is composed of material originating from NGC~4634. We cannot determine whether this material has been ejected tidally or through other processes, which  makes the system highly interesting for follow up observations.}
   {}


\titlerunning{Star-forming dwarf galaxy candidate in  halo of NGC 4634}
\authorrunning{Stein et al.}

\maketitle
%________________________________________________________________
\section{Introduction}
Star formation in spiral galaxies usually takes place close to the midplane of the disk. Nevertheless, star formation high above the midplane is detected in the Milky Way and in external galaxies. Extraplanar H{\sc ii} regions indicate an exchange of matter between the disk and the halo, for example NGC~3628 and the Virgo cluster galaxy NGC~4522 \citep{steinetal2017} as well as NGC~4402 \citep{corteseetal2004}. Some extraplanar H{\sc ii} regions are related to the intracluster or intergalactic medium of interacting galaxies. These H{\sc ii} regions probably formed  in tidal debris \citep{mendesetal2004, ryanweberetal2004} or by ram pressure stripping, for example in the Virgo cluster \citep{gerhardetal2002,oosterloovangorkom2005}. Understanding the physical processes in interacting systems, groups, or clusters is essential to understanding galaxy evolution.

Additionally, dwarf galaxies are apparent in external galaxy groups \citep[e.g.,][]{mulleretal2015} and  in the Local Group \citep[e.g.,][]{mateo1998}. Based on their optical appearance, dwarf galaxies are classified into different categories. Mainly, they are divided into star-forming dwarf galaxies with gas, which are interesting to study as analogs for galaxies in the early Universe \citep[e.g.,][]{vaduvescumccallricher2007}; intermediate types with little star formation and less gas \cite[e.g.,][]{sandagehoffman1991}; and non-star-forming dwarf galaxies with just a little gas, which appear like small elliptical galaxies \cite[e.g.,][]{binggeli94}.
Furthermore, stellar/gaseous tidal debris might recondense within the halo of a merger and form a tidal dwarf galaxy (TDG), e.g., due to local gravitational instabilities in the gaseous component \citep{ducetal2000}. As they are born in situ, their stellar populations are young in comparison to those in normal dwarf galaxies. TDGs can serve as a laboratory for star formation studies in  low-density environments \citep{boquienetal2009}. The building material of TDGs used to belong to a larger parent galaxy and is chemically pre-enriched  compared to normal dwarf galaxies \citep{Duc2012}; i.e., in comparison to dwarf galaxies, TDGs do not follow the luminosity--metallicity relation (L-Z relation). 

Here, we investigate the case of a dwarf galaxy candidate detected by \citet{rossaetal2008} close to the edge-on spiral galaxy NGC~4634. It is located at a projected distance of 1.4 kpc from the NGC~4634 midplane. NGC~4634 is a Virgo Cluster member with enhanced star-forming activity and a neighboring galaxy of NGC~4633. NGC~4634 has a distance of 19.1 Mpc \citep{teerikorpi1992} and it is 2.6\arcmin $\times$ 0.7\arcmin \ in size. The dynamical galaxy mass is 2.7 $\times$ 10$^{10}$~M$_\odot$ \citep{rossaetal2008}. Its H{\sc ii} regions in the disk are widely spread throughout the midplane. This could be due to the interaction with NGC~4633 as the two galaxies are a close binary pair \citep{rotv46341993, rossadettmar2000A&A}. The halo shows diffuse X-ray emission \citep{tuellmann2006} and has a bright diffuse ionized gas component \citep{rossadettmar2000A&A}.\\
\indent NGC~4634 was part of a large sample of H$\alpha$ observations performed by \citet{rossadettmar2000A&A}, who searched systematically for extraplanar diffuse ionized gas (DIG). Among several extraplanar H{\sc ii} regions an extraordinary emission patch was detected in NGC~4634, which showed, in contrast to the other regions, a distributed stellar population in addition to  higher H$\alpha$ emission.

The authors argued that it could be either part of the DIG or a dwarf galaxy in the halo of NGC~4634. Further study by \citet{rossaetal2008}, which is based on the $R$-band morphology, suggests that this emission patch is a gas-rich dwarf irregular galaxy that has been tidally disrupted by the gravitational forces of NGC~4634. However, this study only used optical imaging, and thus the kinematical state and metallicity of the dwarf galaxy candidate have not been taken into account.\\
\indent We further analyze and investigate the origin of this peculiar object, which we  refer to as the dwarf galaxy candidate of NGC~4634, by using various data. We present optical long-slit spectroscopic observations with  ESO Faint Object Spectrograph and Camera (v.2) (EFOSC2) attached to the ESO 3.6~m Telescope, Hubble Space Telescope (HST) r-band and H$\alpha$ data, and Giant Metrewave Radio Telescope (GMRT) H{\sc i} data. We also use photometry of SDSS and GALEX data.\\
\normalsize
\indent The paper is structured as follows. The data reduction and observational strategy of the different observations are presented in Section~2. The analysis and the results of the optical spectroscopy and the other data are described in Section~3. In Section~4 the results are discussed, and in Section~5  the paper is briefly summarized.  

\begin{figure}
   \centering
   \includegraphics[width=0.45\textwidth]{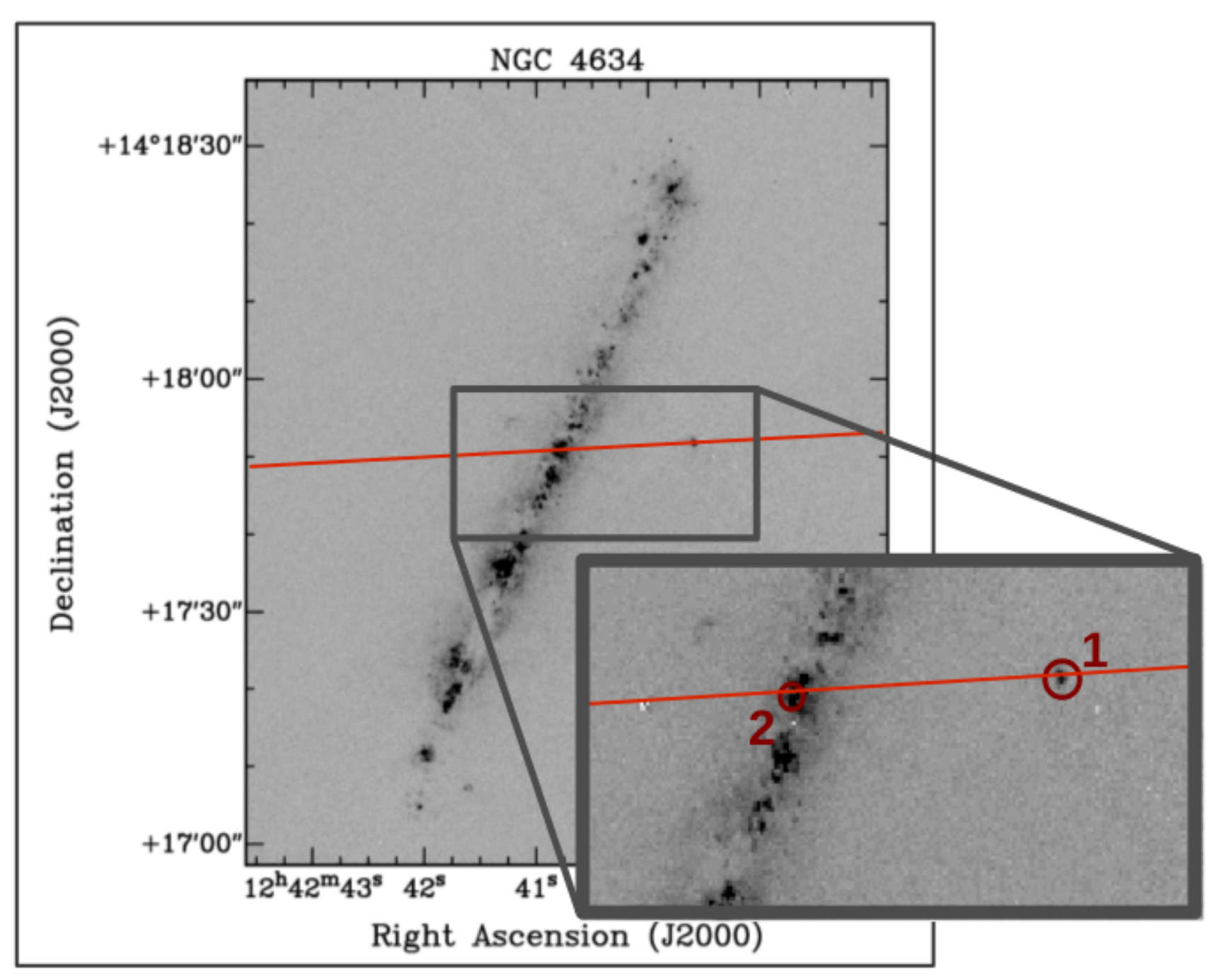}
      \caption{Slit position on the HST H$\alpha$ image of NGC~4634. Indicated in red (see inset) are the regions of interest.}
         \label{NGC 4634}
\end{figure}

%__________________________________________________________________
\section{Data}
\subsection{Long-slit spectroscopic data}
\begin{figure}
\includegraphics[width=\columnwidth]{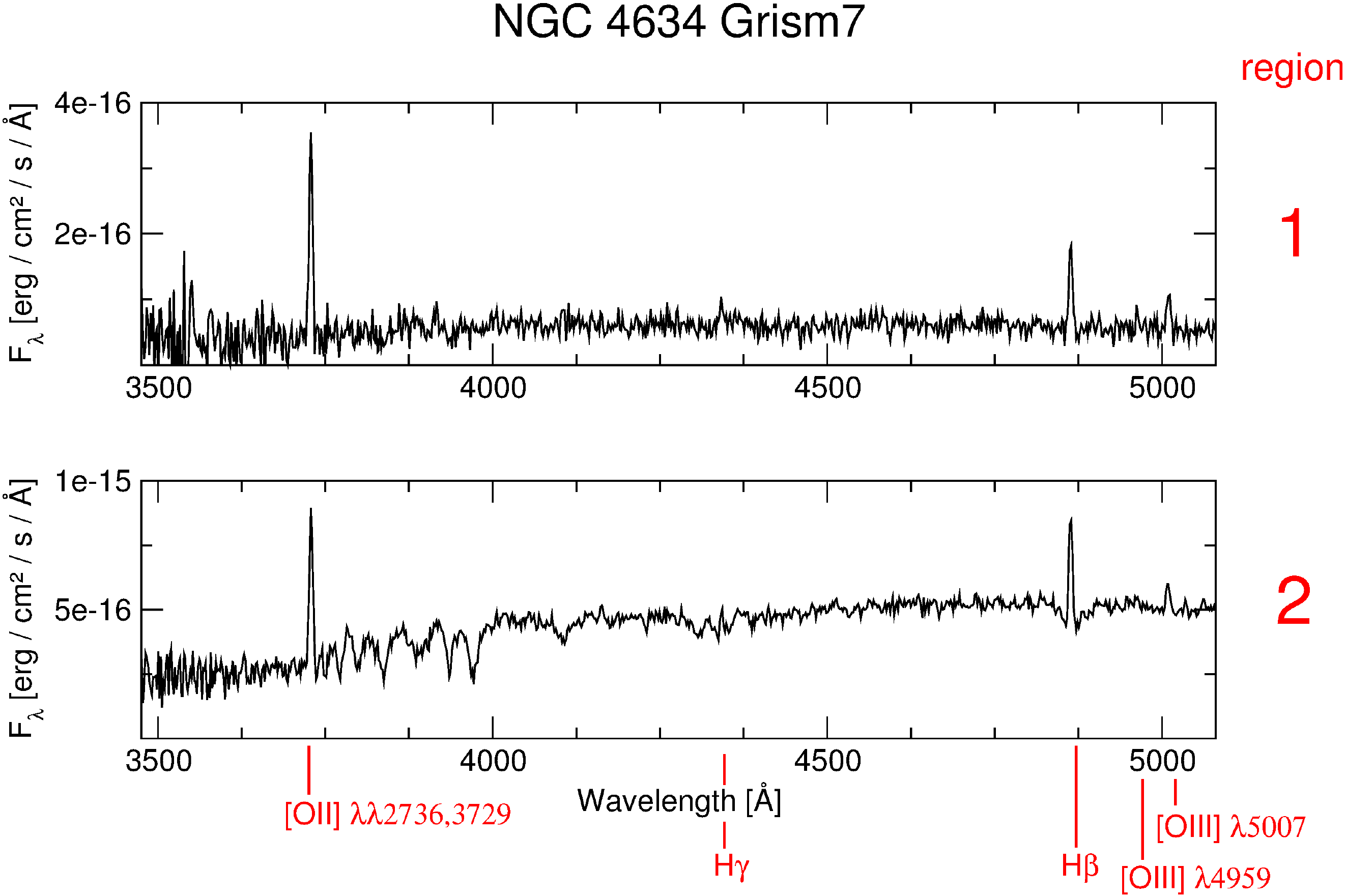}%
\caption{Spectra of grism~7.}%
\label{grism7}%
\end{figure}

\begin{figure}%
\includegraphics[width=\columnwidth]{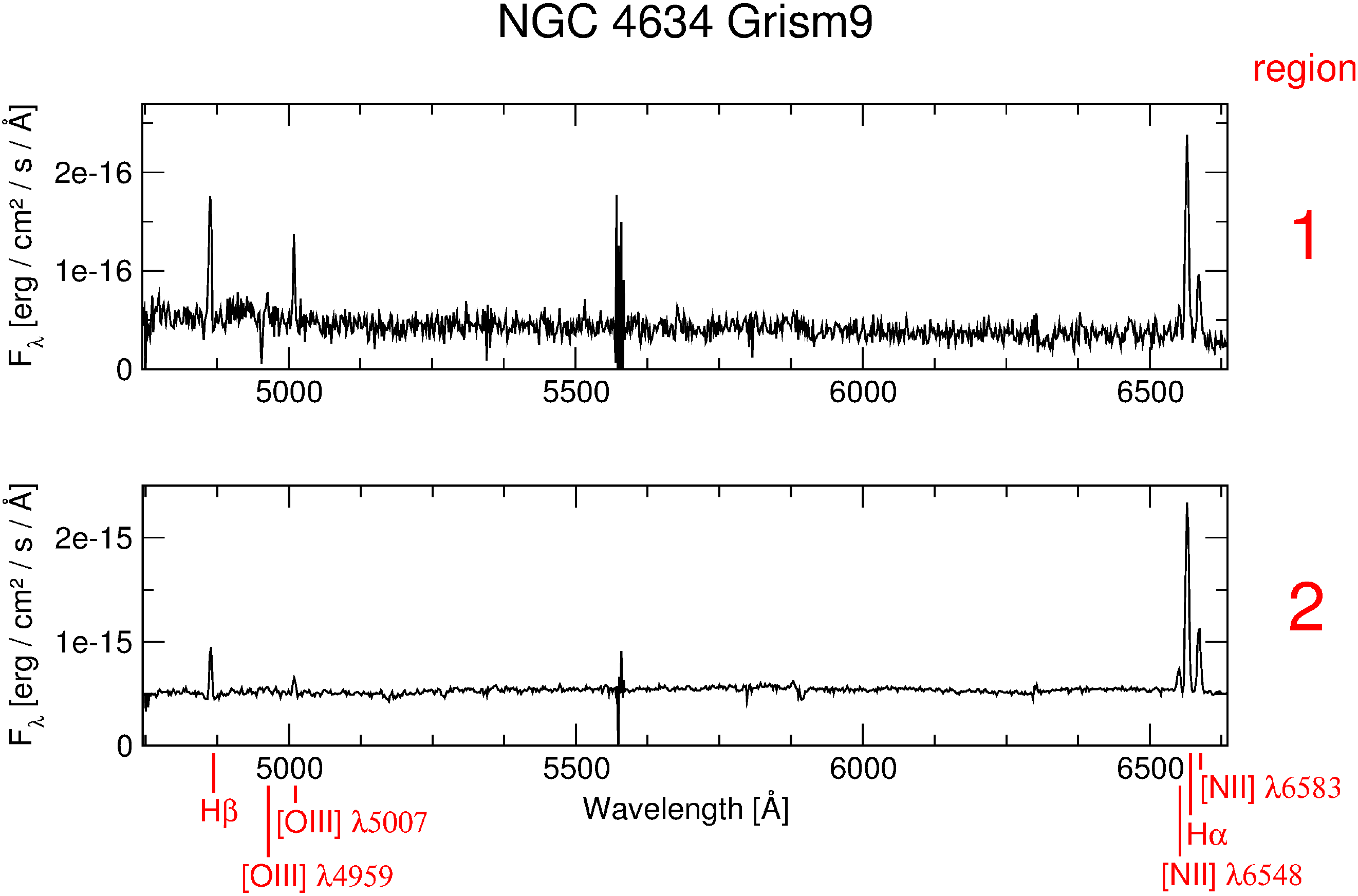}%
\caption{Spectra of grism~9.}%
\label{grism9}%
\end{figure}

The spectroscopic data were obtained at the ESO La Silla observatory with EFOSC2  attached to the 3.6~m Telescope during the nights of 7 -- 9 March 2000. A 2048 x 2048 CCD detector with a pixel size of 15 $\mu$m was used. The field of view is 5.4\arcmin x 5.4\arcmin with a pixel scale of 0.316\arcsec/pixel. With the two different grisms \#7 (3270 -- 5240 \AA) and \#9 (4700 -- 6700 \AA) %, and \#12 (6015 -- 7000 \AA)  
we were able to detect the important lines in the optical. The slit had a length of 5\arcmin and a width of 1\arcsec. The seeing varied between 0.71\arcsec and 0.74\arcsec (see Table \ref{tab:observations}). The total integration time was split  into two exposures of 20 minutes in each grism (see Table \ref{tab:4Galaxien}). In Fig.~\ref{NGC 4634} the slit position with a position angle of 92$^\circ$ is shown superimposed on an H$\alpha$ image of NGC~4634. \\ 
\indent The data reduction was carried out using the image reduction and analysis facility \citep[IRAF,][]{tody1986, tody1993}. This includes the corrections of bias and overscan, response, and cosmic rays with L.A.Cosmic \citep{vandokkum2001}. Additionally, the wavelength calibration was obtained by observing the available helium-argon lamp and the flux was calibrated by observing the standard stars (EG274 and LTT2415). In order to exclude all emission not originating from the H{\sc ii} regions, night sky background emission close to the regions was averaged and subtracted with the IRAF task {\tt background}. This procedure left some residuals at 5580 \AA \ in the grism 9 spectrum of NGC~4634. To correct later for stellar absorption the spectra were coadded before the continuum subtraction. The resulting spectra of the two regions are shown in Figs. \ref{grism7} and \ref{grism9}. Our spectra contain  strong emission lines like the Balmer lines, [OII] $\lambda$$\lambda$3726, 3729, [OIII] $\lambda$$\lambda$4959, 5007, and [NII] $\lambda$$\lambda$6548, 6583. Due to the spectral resolution the lines of [O II] $\lambda$3726.0 and [O II] $\lambda$3728.8 were blended and therefore we use the sum of the doublet lines. 
%, \ref{grism12}.

 \begin{figure}
   \includegraphics[width=0.45\textwidth]{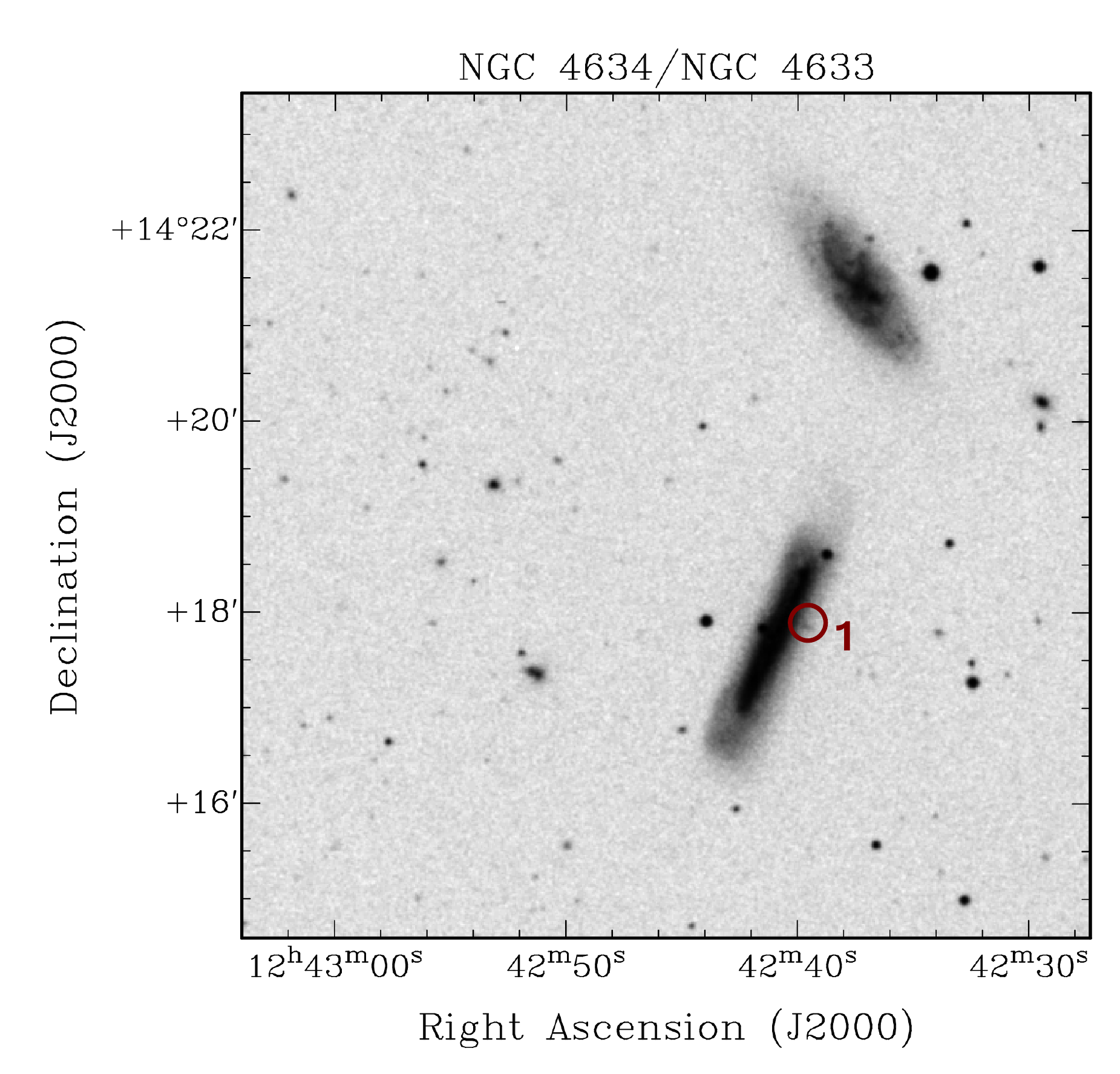}
      \caption{DSS $B$ image of NGC~4634.}
         \label{DSS_NGC 4634}
   \end{figure}
 \normalsize

\begin{table*}
\centering
\begin{threeparttable}
\caption{Basic parameters of NGC~4634}
\begin{tabular}[ht]{ccccccccccc}
\hline\hline 
Galaxy & R.A. (J2000)            & Dec (J2000)           & Hubble type  &      $v$ [km s$^{-1}$]  &  D  [Mpc] & i [$^\circ$]  & PA  [$^\circ$] & Size & Galaxy mass [M$_\odot$]\\
         &    [1]                 &     [2]               &   [3]        &   [4]        & [5]       &     [6]                 &  [7]  & [8] &[9]  \\ [0.2em]  \hline
NGC~4634 &12$^h$42$^m$40.9$^s$\tablefootmark{(1)} &14$^{\circ}$17'45\arcsec\tablefootmark{(1)}& SBcd\tablefootmark{(2)} &  221\tablefootmark{(3)} & 19.1$\pm$0.2\tablefootmark{(4)}&                 83\tablefootmark{(5)} & 155 & 2.6' $\times$ 0.7'\tablefootmark{(3)}&2.7 $\times$ 10$^{10}$\tablefootmark{(3)} \\
         \hline
\end{tabular}
\centering
\tiny
\begin{tablenotes}
\item \textbf{Notes.} [1] Right ascension  and [2] Declination, [3] Hubble type, [4] Radial velocity, [5] Distance, [6] Inclination, [7] Position angle, [8] Major axis $\times$ minor axis, and [9] Total galaxy mass.
\item \textbf{References.}\\
\tablefoottext{1}{NASA/IPAC Extragalactic Database (NED, https://ned.ipac.caltech.edu)}, \tablefoottext{2}{\cite{vau91}},
\tablefoottext{3}{\citet{rossaetal2008}}, \tablefoottext{4}{\citet{teerikorpi1992}}, \tablefoottext{5}{\citet{tuellmann2006}}
\end{tablenotes}
\label{tab:4Galaxien}
\end{threeparttable}
\end{table*}

\begin{table*}
\centering
\begin{threeparttable}
\caption{Observational data}
\begin{tabular}[ht]{ccccccc}
\hline\hline 
                                                                                                                                                         & Filter  & Wavelength                  & t$_{int}$ [s] & Seeing [\arcsec] & AM  & PA [$^\circ$]\\
                                                                                                                                         &[1]      & [2]                         & [3]                             & [4]   & [5]           & [6]\\ [0.2em]  \hline
Long-slit spectroscopy ESO 3.6 m Telescope       & grism 7   & 3270 -- 5240 \AA     & 2x1200      &  0.74  & 1.41 &  92  \\
                                                                                                                                  & grism 9              &       4700 -- 6700 \AA     &       2x1200       &  0.71    & 1.40 & 92  \\
    Hubble Space Telescope                                                      & H$\alpha$               &       6562.82 \AA                     &       6888         &       -       & -  &\\
    Hubble Space Telescope                                                      & r-band  &       5446 -- 7100            &               2308        &           -       & - & \\
    Giant Meter Radio Telescope                                         & HI                  & 21 cm                                             &19700     &           -       & -&  \\
         \hline
\end{tabular}
\centering
\tiny
\begin{tablenotes}
\item \textbf{Notes.} [1] Filter, [2] Wavelength range or central wavelength, [3] Integration time, [4] Seeing, [5] Airmass, and [6] Position angle of the long slit.
\end{tablenotes}
\label{tab:observations}
\end{threeparttable}
\end{table*}

\normalsize

\subsection{HST imaging}
NGC~4634 was observed with the Advanced Camera for Surveys (ACS) at HST as part of program \#10416 (PI: Rossa) in the HST-ACS r-band (F625W) and the H$\alpha$ filter (F658N) using the the Wide Field
Channel (WFC) for 2308s and 6888s, respectively. Earlier results
from the program were reported by \citet{rossaetal2008}, \citet{Dettmar2010}, and \citet{Rossa2012}. 
We retrieved the imaging data for NGC~4634 from the Space Telescope Science Institute (STScI) HST archive, using  
on-the-fly reprocessing using the latest calibration files. The pipeline included the 
new, improved treatment of charge transfer efficiency (CTE) correction, and drizzling the subexposures to cosmic-ray cleaned, 
astrometrically improved image, as discussed in the ACS Data handbook 
V9.0 \citep{Lucas2018}. The pipeline processed and dizzled final images of both filters F625W and F658N were then aligned to subpixel accuracy using IRAF tasks (imexamine, imcentroid, imshift) with several stars in the ACS field. The 
F625W image is used as a continuum image to correct the H$\alpha$ emission in 
the F658N filter for its residual continuum. 
The fluxes of the Galactic foreground stars were used
to determine an empirical scaling factor between the F625W and
F658N fluxes, which we used to rescale the continuum image to the line plus 
continuum image. Subtracting the thus scaled F625W image from the F658N image
yields a pure narrow line image representing the H$\alpha$ emission plus some contribution
from the [NII] lines, which is also included in the filter band at the given redshift of NGC~4634. 
Due to the possibility of a slight difference
of the continuum slope of the Galactic foreground stars and the
continuum emission of the stellar disk, we interactively changed
the scaling factor around the starting value given by the Galactic
stars in order to find the best value. This best value was considered
to be the scaling factor that removed the stellar continuum emission of the disk of NGC~4634
from the F658N image without creating regions of negative residuals.
It was determined by optically inspecting the corrected image.
The optimal value of the scaling factor turned out to be only 2 \%\ lower than the formal 
scaling factor derived from the foreground stars, showing that the above-mentioned effect of 
the color difference between the foreground stars and the integrated color of the stellar 
disk of NGC~4634 is very small. The uncertainty of the scaling factor is even
below the value quoted above since applying the ``formal'' value still left
noticeable continuum artifacts in the disk. The uncertainty of the scaling
parameter (and therefore the stellar continuum flux) is in this case less important than
the intrinsic calibration uncertainty using the broadband  image, as discussed
for ground-based images by, e.g., \citet{GildePaz2003}.

Flux calibration of the narrowband image was then performed using the PHOTFLAM keyword, as outlined in the ACS data handbook V9.0 \citep{Lucas2018}. 
Using a broadband image for the continuum subtraction is not ideal since the line emission contributes weakly  to the
broadband filter, but it is a standard method that still yields  well-calibrated line images 
\citep[see, e.g.,][]{GildePaz2003}.
Photometry of the resolved sources was done with DAOPHOT using the aperture
photometry task and directly calibrated to point sources with photometry in the HST Source Catalog V1 \citep{Whitmore2016}.

\subsection{GMRT data}
In May 2013 NGC~4634 and its companions were observed for 5.5 hrs with the GMRT. To calibrate the flux and the bandpass, 3C286 was observed for 15 min at the start and the end of the run and 1264+116 was observed every 30 min in order to calibrate the phases throughout the run. The correlator was set up with 4 MHz bandwidth and 512 channels centered at 1419.3 MHz.\\
\indent The data were reduced with a range of data reduction packages. The raw data, in lta format, is first converted into a fits file through the use of listscan (v2.0) and gvfits (v2.0) as provided by the observatory. The produced fits file is then loaded into the data reduction package {\sc MIRIAD} \citep{Sault1995} where RFI is flagged in an automated manner. The data are then further flagged and calibrated through the use of the package {\sc FLAGCAL} \citep[v0.989,][]{Chengalur2013}. After this initial calibration the data is loaded into {\sc CASA} \citep{McMullin2007}. We first visually inspect all the baselines for any residual RFI and then self-calibrate the data through the standard {\sc CASA} procedures. After the self-calibration we subtract the continuum in the $uv$-plane from the dataset with {\sc UVCONTSUB} by fitting polynomials of order one. The 8.1 kHz channels were combined into 24.3 kHz channels with the task {\sc SPLIT}, and {\sc TCLEAN} was used to transform the visibilities into a data cube. The visibilities were weighted according to a Briggs weighting scheme with Robust~=~0. The dirty cubes were cleaned  with the {\sc TCLEAN} multiscale clean algorithm, at scales corresponding to 1 and 4 beams. The cleaning was performed in an iterative process where we first cleaned the cube to a threshold of 6$\sigma$, created a mask with SoFiA \citep{Serra2015} and then cleaned within this mask to 0.5$\sigma$. This procedure was followed until successive iterations showed no changes, while all the emission visible was captured in the mask.\\
\indent We constructed different cubes, one with a uv taper of 7 k$\lambda$ and one with a taper of 20 k$\lambda$. This resulted in two cubes with a channel width of 5.2 km s$^{-1}$, one with a Gaussian beam FWHM of 18\arcsec$\times$17\arcsec and one with a resolution of FWHM = 6\arcsec$\times$5\arcsec. Additionally, a Hanning smoothed version of the high-resolution cube with a velocity resolution of 10.4 km s$^{-1}$ was created. 

%_______________________________________________________________________________________________________________________________
 
 \section{Analysis and results}
 \subsection{Long-slit spectra}
 \subsubsection{Emission line measurement}
We analyzed two H{\sc ii} regions from the long-slit spectra (see Fig.~\ref{NGC 4634}). The first region is called the H{\sc ii} region (1); it is in the halo of NGC~4634, which is associated with the dwarf galaxy candidate.  The second  region is called the disk H{\sc ii} region (2), and lies in the disk of NGC~4634.\\
\indent The spectra (Fig. \ref{grism7}, Fig. \ref{grism9}) were analyzed via the {\tt splot} task of IRAF. 
We chose the size of the extraction area of the dwarf galaxy candidate's H{\sc ii} region (1) and disk H{\sc ii} region (2) of NGC~4634 depending on the extent of their H$\alpha$ emission along the spatial axis in the spectra. The identified emission lines are marked red in Fig.~\ref{grism7} and Fig.~\ref{grism9}. The emission line fluxes were measured with the {\tt deblending} tool within the {\tt splot} task of IRAF. With the Gaussian fit function we measured fluxes and central peak positions. 
 
 \subsubsection{Stellar absorption correction and extinction correction}
\begin{figure*}
\begin{center}
\sidecaption{\resizebox{11cm}{!}{\includegraphics{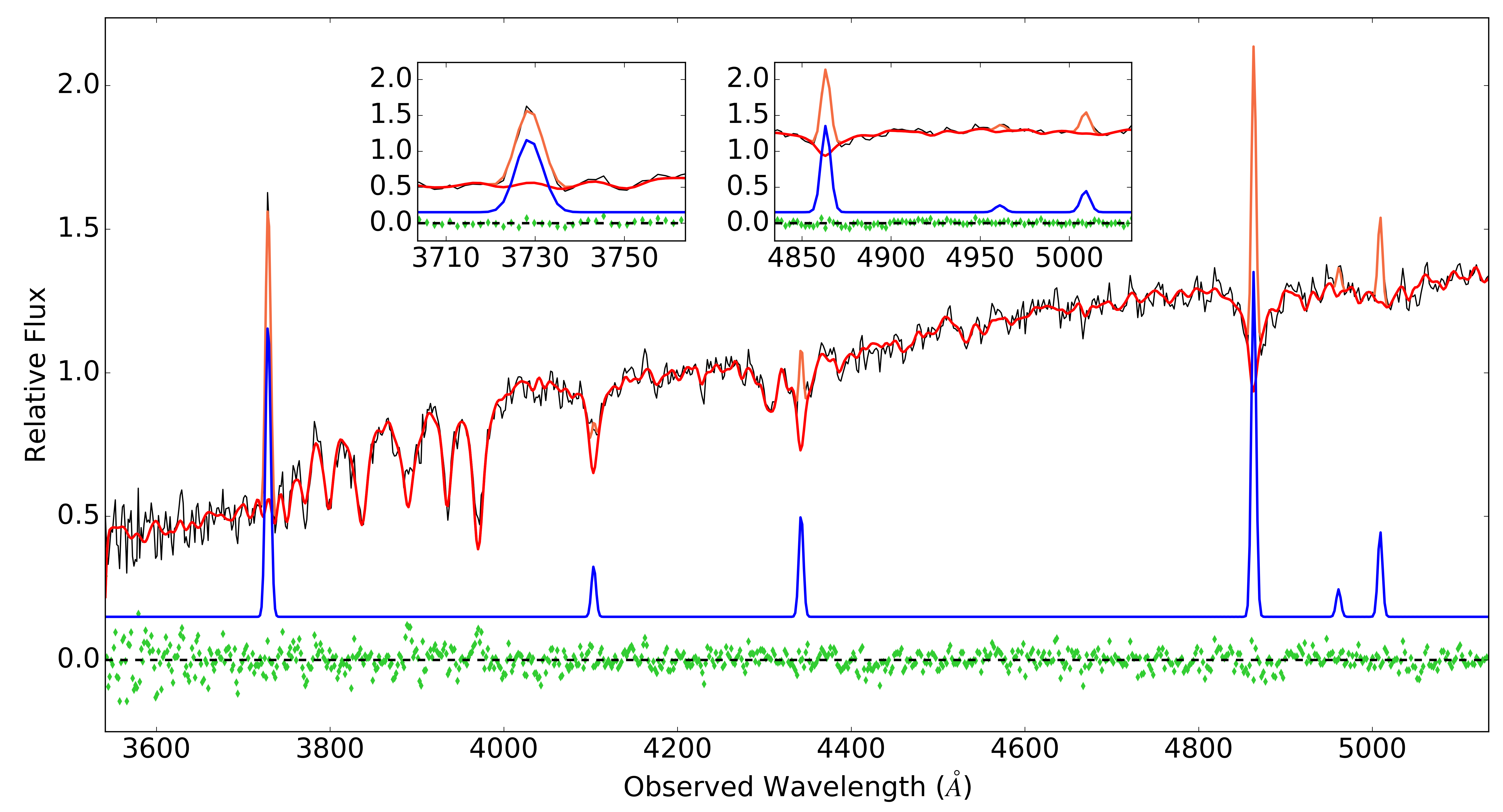}}}
\caption{pPXF spectrum of disk H{\sc ii} region (2) in grism~7. The black line represents the observed spectrum; the orange line shows the emission lines fitted by pPXF to the observed spectrum. The blue line is the extracted pure emission line spectrum; the red line is the fitted continuum to the data; the green dots are the residuals around the zero value (gray dashed line).}
        \label{fig:ppxfNGC4634-gr7-r2}
\end{center}
\end{figure*}

\begin{figure*}
\begin{center}
\sidecaption{\resizebox{11cm}{!}{\includegraphics{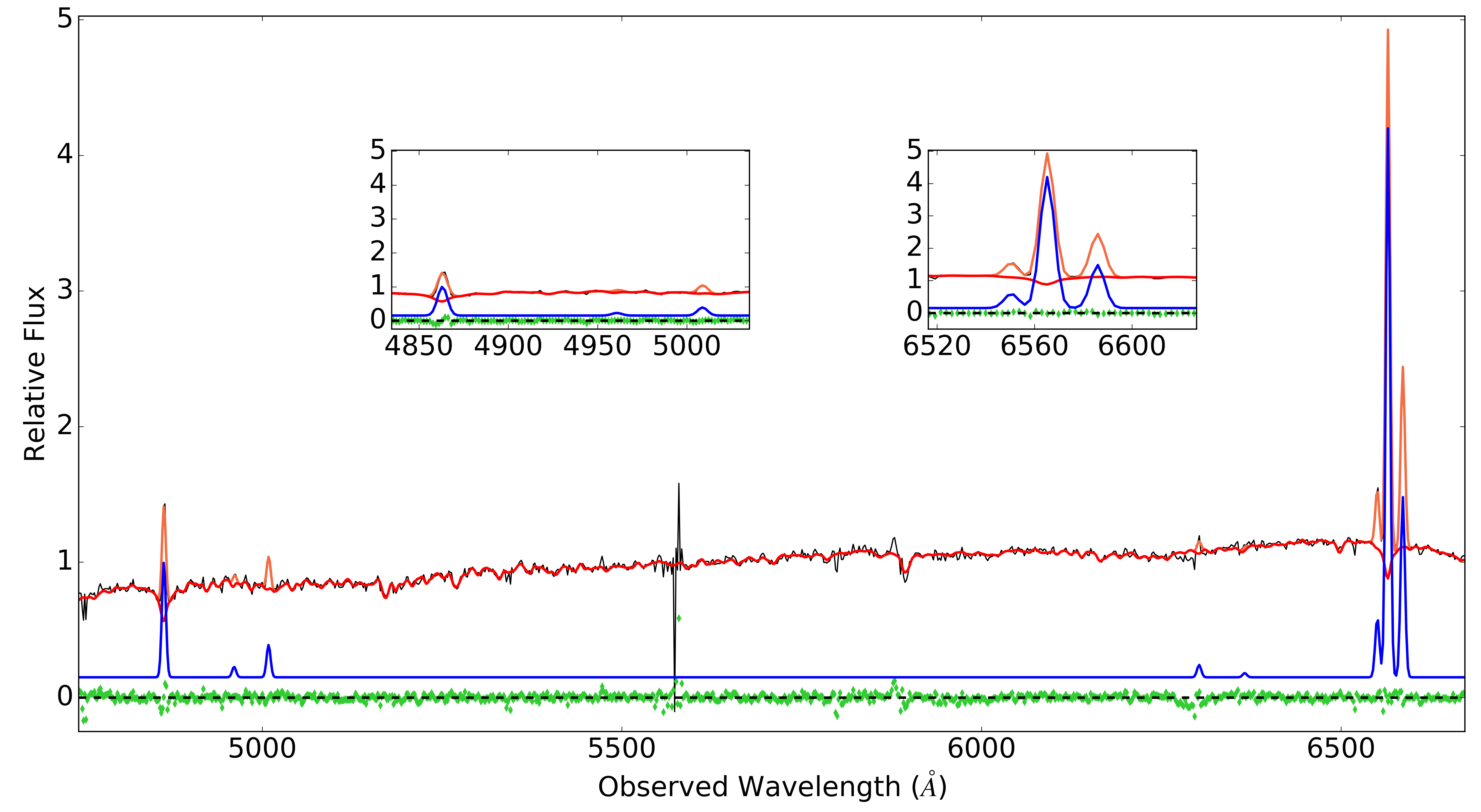}}}
\caption{pPXF spectrum of disk H{\sc ii} region (2) in grism~9. The black line represents the observed spectrum; the orange line shows the emission lines fitted by pPXF to the observed spectrum. The blue line is the extracted pure emission line spectrum; the red line is the fitted continuum to the data; the green dots are the residuals around the zero value (gray dashed line).}
        \label{fig:ppxfNGC4634-gr9-r2}
\end{center}
\end{figure*}

Absorption lines of the underlying stellar populations can have an influence on the measurement of emission lines, especially the Balmer lines. Therefore, we investigated this effect in our spectra. We identified strong underlying stellar continuum including absorption in the disk H{\sc ii} region (2) of NGC~4634. A little stellar continuum is visible in the dwarf galaxy candidate's H{\sc ii}~region~(1). To correct for this effect we adopted pPXF fitting \citep[][upgrade]{Capellari2004, Cappellari2017} (see Fig.~\ref{fig:ppxfNGC4634-gr7-r2} and Fig.~\ref{fig:ppxfNGC4634-gr9-r2}). To account for internal extinction the absorption corrected fluxes of the disk H{\sc ii} region~(2) were dereddened with the IRAF task {\tt redcorr}. For this we used the theoretical Balmer decrement of H$\alpha$/H$\beta$ = 2.86 (Case B, 10\,000 K) and the Galactic extinction law \citep{savagemathis1979}. In the dwarf galaxy candidate the measured flux ratio of H$\alpha$ to H$\beta$ is 2.86. As this is equal to the theoretical Balmer decrement we did not correct for absorption and did not deredden H{\sc ii} region~(1). The H$\beta$ fluxes and the dereddened scaled fluxes relative to H$\beta$ are listed in Table \ref{table:Hbeta}.
 
\begin{table}
\caption{Results from optical long-slit spectroscopy, magnitudes from different filters and oxygen abundances of the dwarf galaxy candidate~(1) and one H{\sc ii} region of NGC~4634~(2).}             
\label{table:2} 
\label{table:Hbeta} 
\label{parameter}  
\centering          
\begin{tabular}{l c c}    
\hline\hline       
Region                                                                                  &  1           &   2     \\ 
\hline
H$\beta$ $[$ 10$^{-16}$ erg cm$^{-2}$ s$^{-1}$] & 6.53 $\pm$ 0.94 & 2.07 $\pm$ 0.70 \\
$[$OII$]$ $\lambda\lambda$ 3726, 3728 / H$\beta$ & 2.38 $\pm$ 0.45      & 1.27 $\pm$ 0.24\\
$[$OIII$]$ $\lambda$4959  / H$\beta$    & 0.15 $\pm$ 0.04 & -  \\
$[$OIII$]$ $\lambda$5007  / H$\beta$    & 0.60 $\pm$ 0.10                & 0.35 $\pm$ 0.06 \\
$[$NII$]$  / H$\beta$                           &       0.45 $\pm$ 0.07   & 0.30 $\pm$ 0.05  \\
H$\alpha$       / H$\beta$                                      &       2.86 $\pm$ 0.31        & 2.86 $\pm$ 0.31  \\
$[$NII$]$ / H$\beta$                                    &       1.13 $\pm$ 0.17      & 1.00 $\pm$ 0.15  \\ 
c                                                                        & 0.00                  & 0.38  \\ 
v$_{helio}$ [km s$^{-1}$]    & 166 $\pm$ 23                     & 161 $\pm$ 23      \\
M$_B$                                                    &      -10.76 $\pm$ 0.15    &                       \\
M$_R$                                                                                    &       -10.87 $\pm$ 0.15       &                       \\
SDSS u' [mag]                                                                            &21.01 $\pm$ 0.10  &                    \\
SDSS g' [mag]                                                                            &20.82 $\pm$ 0.12 &                     \\
SDSS r' [mag]                                                                            &20.66 $\pm$ 0.15 &                     \\
GALEX FUV [mag]                                                                          &19.67 $\pm$ 0.25 &                     \\
GALEX NUV [mag]                                                                          &19.91 $\pm$ 0.20 &                     \\
UKIDSS K [mag]                                                                          & $<$~20.7    &                           \\
log(L(H$\alpha$)) [erg s$^{-1}$]  & 38.88       $\pm$ 0.16 &                                    \\
log(SFR$_{\text{H}\alpha}$)                     & -2.33 $\pm$ 0.13 &         \\
\hline
\textbf{12~$+$~log(O/H)}                                                        &       \                         &       \       \\
\citet{kewleydopita2002}    & 8.99 $\pm$ 0.04 & 9.08 $\pm$ 0.04         \\
\citet{pilyuginetal2014}    & 8.51 $\pm$ 0.13 & 8.62 $\pm$ 0.11    \\
\citet{pettinipagel2004}    & 8.67 $\pm$ 0.16 & 8.73 $\pm$ 0.15    \\
mean                                                                                            &       8.72  $\pm$ 0.11&    8.81 $\pm$ 0.10         \\
\hline                  
\end{tabular}
\end{table} 
 
\noindent
\subsubsection{Heliocentric velocities} 
The velocities of both H{\sc ii} regions were determined with the redshifts of the [OIII] $\lambda$5007 line, which was bright and closest to the center of grism 9 where the wavelength calibration is most reliable. The heliocentric correction was calculated by the IRAF task {\tt rvcorrect}.\\
\indent The resulting heliocetric velocity (Table~\ref{parameter}) of the dwarf galaxy candidate's H{\sc ii} region (1) is similar to that of the H{\sc ii} region (2) of the galaxy,  around 160 km s$^{-1}$. Comparing the determined heliocentric velocities with our velocity map heliocentric velocities we find that they are consistent with those derived from the H{\sc i} (Fig.~\ref{Velmap}, see Section~3.5).

\subsection{HST r-band and continuum-corrected H$\alpha$ images}
\begin{figure}[ht]
        \centering
                \includegraphics[width=0.45\textwidth]{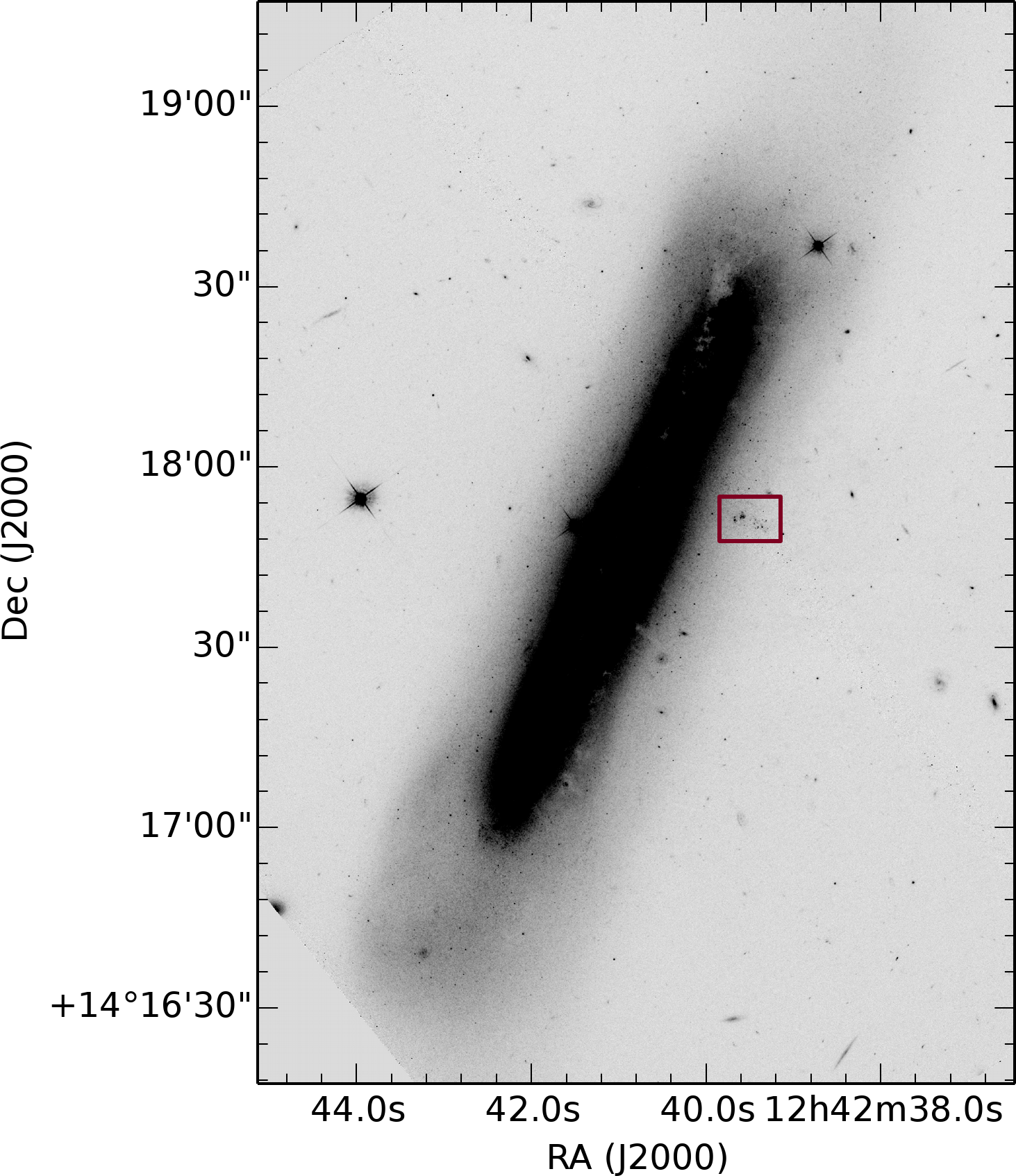}
        \caption{NGC~4634 F625W image, from HST/ACS, the box shows the region of Fig. \ref{fig:NGC4634-dwarf-ha} and Fig. \ref{fig:NGC4634-dwarf-r}.}
        \label{fig:NGC4634-HST}
\end{figure}

\begin{figure}[ht]
        \centering
                \includegraphics[width=0.45\textwidth]{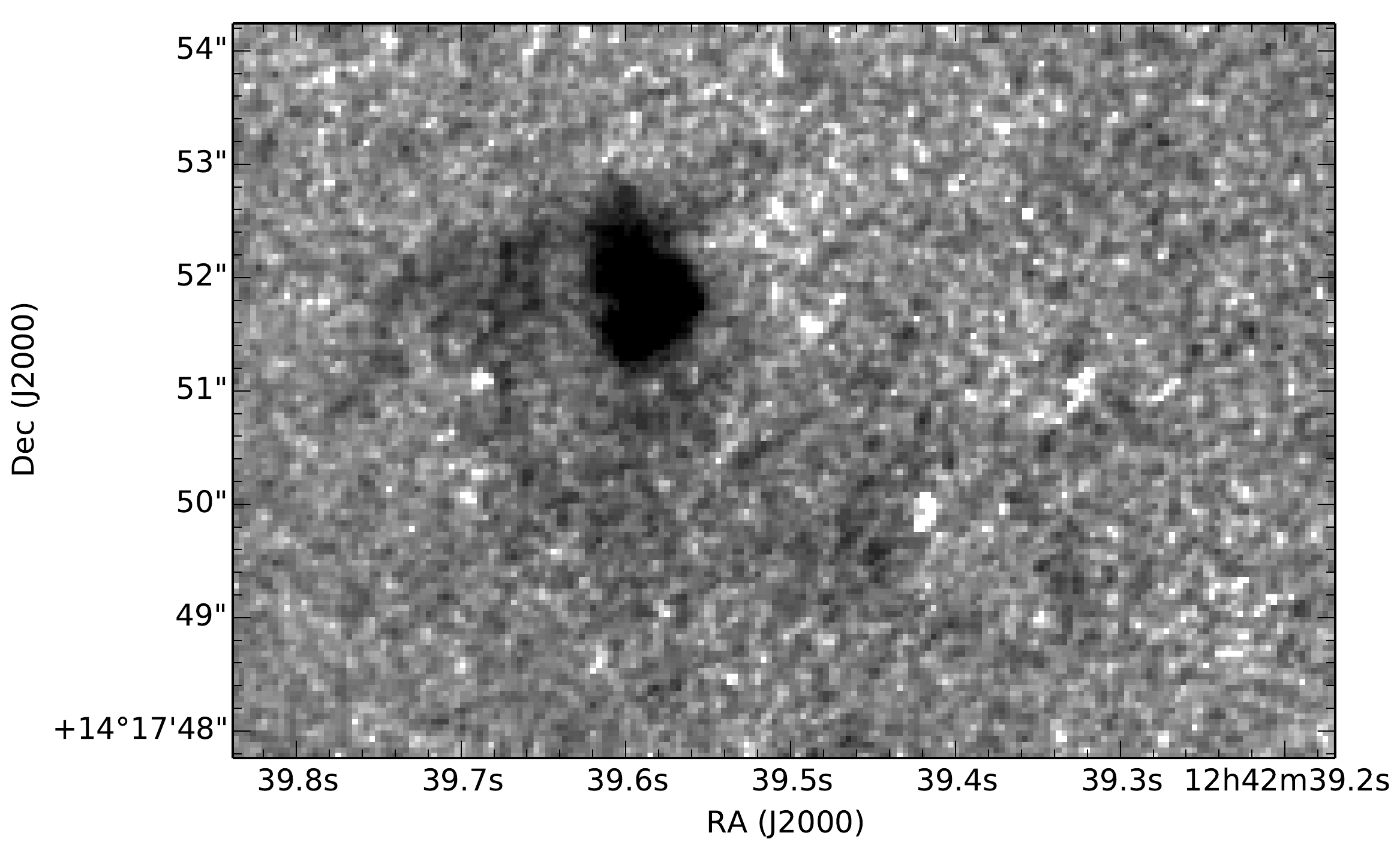}
        \caption{H$\alpha$ image of the dwarf galaxy candidate.}
        \label{fig:NGC4634-dwarf-ha}
\end{figure}

\begin{figure}[h!]
        \centering
                \includegraphics[width=0.45\textwidth]{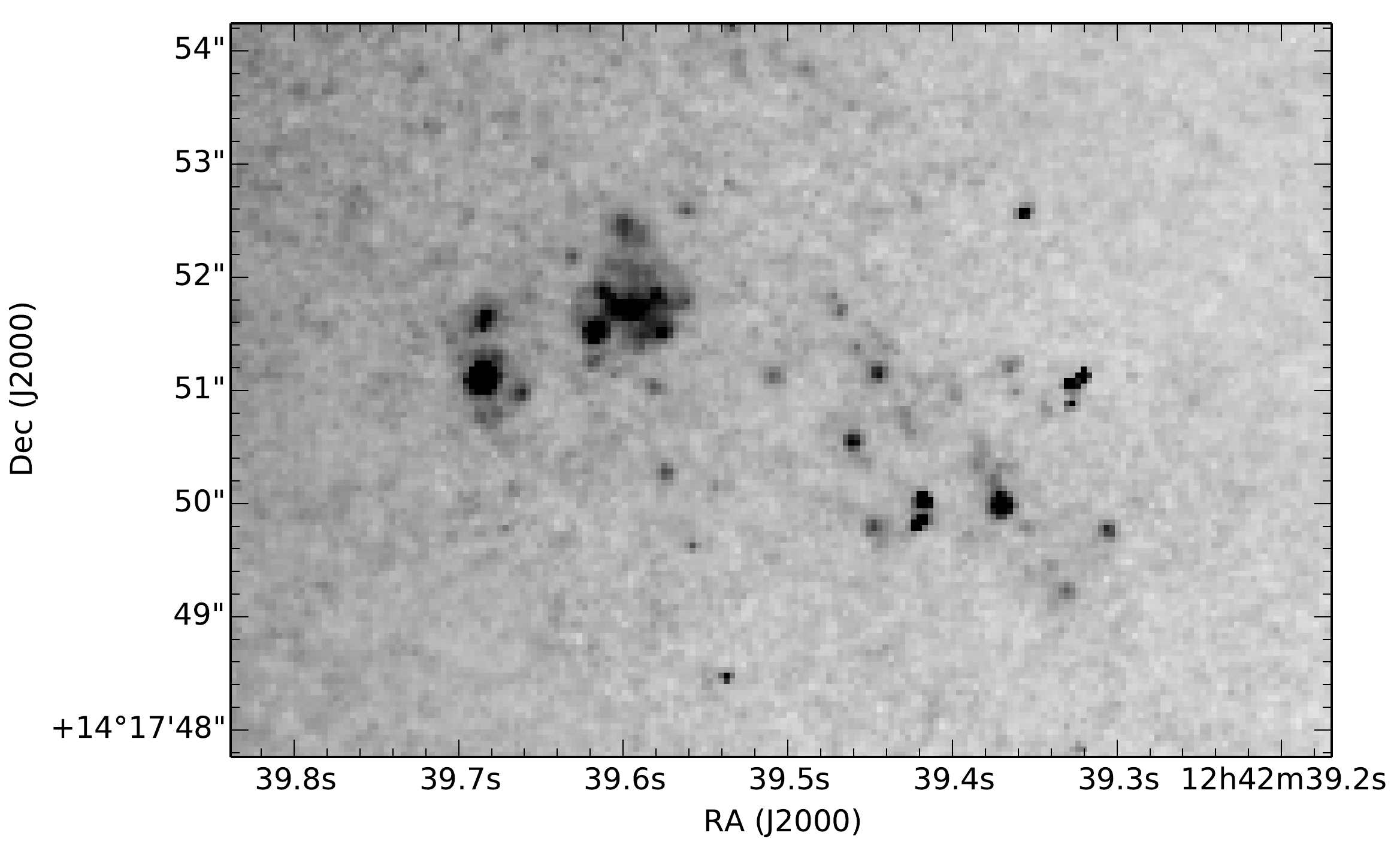}
        \caption{F625W image of the dwarf galaxy candidate.}
        \label{fig:NGC4634-dwarf-r}
\end{figure}

With the r-band (F625W) and corrected H$\alpha$ images (Figs. \ref{fig:NGC4634-HST}, \ref{fig:NGC4634-dwarf-ha}, \ref{fig:NGC4634-dwarf-r}) observed by the HST we are able to analyze the galaxy and the dwarf galaxy candidate visually.\\
\indent The HST F625W image of NGC~4634 (Fig. \ref{fig:NGC4634-HST}) shows  an elongation from the major axis, especially to the south and north. This is probably due to tidal interaction with NGC~4633. The galaxy pair is shown in Fig. \ref{DSS_NGC 4634}. \\
\indent Figure \ref{fig:NGC4634-dwarf-ha} shows the subarea of the ACS image centered on the dwarf galaxy candidate. Most of the H$\alpha$ emission originates from the H{\sc ii} region (1) to the east. This H{\sc ii} region seems to have a bubble structure. Nevertheless, there is diffuse H$\alpha$ spread over the whole dwarf galaxy candidate.
The dwarf galaxy candidate has a projected distance to the plane of 1.4 kpc at the distance of NGC~4634. This was determined as the distance from the galaxy's midplane (defined by the central line from the H$\alpha$ image) to the dwarf galaxy candidate's H{\sc ii} region (1). Its main H$\alpha$ emission in Fig.~\ref{fig:NGC4634-dwarf-ha} has a diameter of 1\arcsec~$\approx$~90 pc, including the diffuse emission and the stars the diameter increases to 6\arcsec~$\approx$~550 pc.
The F625W image (Fig. \ref{fig:NGC4634-dwarf-r}) shows an elongated body pointing roughly perpendicular toward the disk of NGC~4634. The continuum image is partly resolved into point sources, which are especially visible at the location of the bubble-like H{\sc ii} emission region (see Fig.~\ref{fig:NGC4634-dwarf-ha}). The stars to the southwest are not foreground stars but can clearly be related to the dwarf galaxy candidate. 


\subsection{H$\alpha$-luminosity and star formation of the dwarf galaxy candidate}
The integrated H$\alpha$-flux (I$_{\text{H}\alpha}$) of the dwarf galaxy candidate's H{\sc ii} region (1) was determined via aperture photometry on the H$\alpha$-image (Fig. \ref{fig:NGC4634-dwarf-ha}). With I$_{\text{H}\alpha}$~=~(1.7~$\pm$~0.5)~$\times$~10$^{-14}$~erg~s$^{-1}$~cm$^{-2}$ we computed the luminosity and the number of photons produced per second following the Osterbrock (1989) case B with a temperature  of 10$^4$ K:
\begin{align}
\text{L}_{\text{H}\alpha} = (7.5 \pm 2.2) \times 10^{38} \text{erg s}^{-1}\\
Q(\text{H}^0,\:T)_{\text{H}\alpha} \ = (2.7 \pm 0.8) \times 10^{50} \text{photons s}^{-1}
\end{align}
To calculate the Lyman-$\alpha$ photons from this we multiply by a factor of 2.2 \citep{hummerstorey87}. The final number of photons is then 
\begin{align}
Q(\text{H}^0,\:T)_{\text{Ly}\alpha} \ = (6.0 \pm 1.8) \times 10^{50} \text{photons s}^{-1}.
\end{align}
According to \citet{panagia73} several O4 stars or hundreds of B0 stars are able to produce this amount of photons.
Additional stars around the H{\sc ii} region are visible in the $R$-band image (Fig. \ref{fig:NGC4634-dwarf-r}).\\
\indent We determined the corresponding H$\alpha$ star formation rate from \citet{kennicutt1998}:
\begin{align}
\text{SFR} = 7.9 \times 10^{-42} \ \text{L}_{\text{H}\alpha} (\text{erg s}^{-1}) = (4.7 \pm 1.4) \times 10^{-3} \text{M}_\odot \text{yr}^{-1}
\end{align}

\subsection{Stellar populations and stellar mass}
\label{subsec:stellar}
\label{subsec:stellarmass}
The dwarf galaxy candidate is not only visible in our HST images, but it is also 
present on the blue SDSS images (filters u', g', and r') of NGC~4634 and the Galaxy Evolution Explorer (GALEX) far-UV (FUV) and near-UV (NUV) images. Interestingly, the dwarf galaxy candidate is absent on the red SDSS images (filters i' and z') and in the near-infrared images from UKIRT Infrared Deep Sky Survey (UKIDSS) in the J, H, and K filters.  A marginal detection in the Y filter coincides with the brightest knot in H$\alpha$ and may
therefore be caused by the emission of the [SIII] $\lambda 9512$ line, which 
is the brightest HII emission line in the Y filter.\\
\indent We performed aperture photometry on the SDSS and GALEX images, through 
apertures with radius of 3.2\arcsec and 5.3\arcsec, respectively, centered on the 
GALEX FUV emission.  
The SDSS photometry was directly calibrated by comparison to the nearest 
isolated star. For the GALEX flux we picked a more distant star to ensure 
a solid measurement in the FUV filter. For the K filter of UKIDSS we derived an upper limit. The derived magnitudes are listed in Table \ref{table:2}.\\
\indent With the distance modulus of m~-~M~=~31.4~mag and assuming no extinction, we 
calculated the absolute magnitudes in $B$-band (M$_B$) and in $R$-band (M$_R$) of the dwarf galaxy candidate using 
the SDSS magnitudes and the conversion relations by Lupton 2005\footnote{https://www.sdss3.org/dr10/algorithms/sdssUBVRITransform.php}, 
which are also useable for galaxies,  according to the notes of the 
SDSS web page. Based on the conversion relation we get M$_B$~=~-10.76 and M$_R$~=~-10.87.
The blue colors (both in the optical and the UV) of the dwarf galaxy candidate reflect its ongoing star formation and hints at a lack of a detectable
intermediate or old stellar population.\\
\indent Using the GALEX fluxes and the calibration of \citet{leeetal2009} we estimated 
the star formation rate from the UV stellar continuum to be 
SFR = 1.0 $\pm 0.1 \times 10^{-3}$ M$_{\odot}$ yr$^{-1}$, a factor of 5 lower than the H$\alpha$ derived
star formation rate. Since the GALEX UV bands also probe
stars somewhat older than hydrogen ionizing stars, this
may imply that the star formation rate of the dwarf galaxy candidate is
currently higher than $\sim$100 Myr ago.\\
\indent Using the equation in \citet{belletal2003} for the SDSS bands from their Table 7, we calculate the mass-to-luminosity ratio for r'-band with the u' and g' filters,
\begin{equation}
log(M/L)_r = -0.99 + (0.345 \cdot (u' - g') = -0.033
,\end{equation}
which leads to $M/L = 0.9 \pm 0.1$. With the absolute magnitude from r-band we get $L=1.5 \times 10^6 L_\odot$, and from that a stellar mass of 1.35~$\times$~10$^6 M_\odot$.

Furthermore, no old stellar population seems to be present down to our detection limits with a 3$\sigma$ level of m$_K$ = 20.7 mag from UKIDSS survey data \citep{Lawrence2007}. 
This implies an upper mass limit for the old stellar population (age~$<$~1~$\times$~10$^9$~yr) of 2.40~$\times$~10$^3 M_\odot$ (based on SED modeling with FAST \citep{krieketal2009}). 
We also note that the ACS F658W (HST r-band) shows only few point sources at the location of the tidal dwarf candidate, consistent with 
being blue supergiants and/or young clusters, and no extended low surface 
brightness, granular body consistent with significant mass in an old stellar population.

\subsection{Oxygen abundance}
To derive oxygen abundances, we used strong-line calibrations. There are different strong-line calibration types, empirical \cite[e.g.,][]{pilyugin2001, pettinipagel2004, pilyuginetal2014} and theoretical \cite[e.g.,][]{kewleydopita2002, KK2004}, which show well-known offsets between 0.1 and 0.7 dex \citep{stasinska2002, modjazkewley2008, moustakas2010}. The metallicities were derived using the oxygen abundance 12~+~log(O/H) by three different methods in order to have comparable results:
\begin{itemize}
\item \cite{kewleydopita2002} (theoretical calibration);
\item \cite{pilyuginetal2014} (empirical calibration); 
\item \cite{pettinipagel2004} (empirical calibration).
\end{itemize} 
For further information, discussion, and the error analysis see \citet{steinetal2017}.

%%%%%%%%%%%%%%%%%%%%%%%%%%%%%%%%%%%%%%%%%%%%%%%%%%%%%%%%%%%%%%%%%%%%%%%%%%%%%%%%%%%%%%%%%%%%%%%%%%%%%%
The results of the abundance determination (Table \ref{parameter}) show slightly lower but  similar (within the errors)\ oxygen abundances of the dwarf galaxy candidate's H{\sc ii} region (1) in comparison to the disk H{\sc ii} region (2) with each calibration.\\
\indent Due to the known and here confirmed offsets between the different calibrations, which are in agreement with other analyses \citep{modjazkewley2008, kewleyellison2008}, we use the average of all three oxygen abundances of 12~+~log(O/H)~=~8.72 for the dwarf galaxy cadidate.

\subsection{Luminosity--metallicity relation (L-Z Relation)} 
In order to determine whether the dwarf galaxy candidate was created from gas stripped from NGC~4634 or a dwarf galaxy in the process of being accreted, we use the well-defined relationship between the absolute $B$-band luminosity and the metallicity of galaxies known as the L-Z relation, which also applies to star-forming dwarf galaxies  \citep[e.g.,][]{richermccall1995}.  %Skillman et al.1989 
This can be applied  to the dwarf galaxy candidate in a straightforward manner. The metallicity was taken from our mean oxygen abundance of the dwarf galaxy candidate of 12~+~log(O/H)~=~8.72. The absolute luminosity in $B$-band (M$_B$) was calculated from SDSS aperture photometry on the u', g', and r' images (see Sect. \ref{subsec:stellar}). 

\begin{figure*}
\begin{center}
\sidecaption
{\resizebox{13.5cm}{!}{\includegraphics{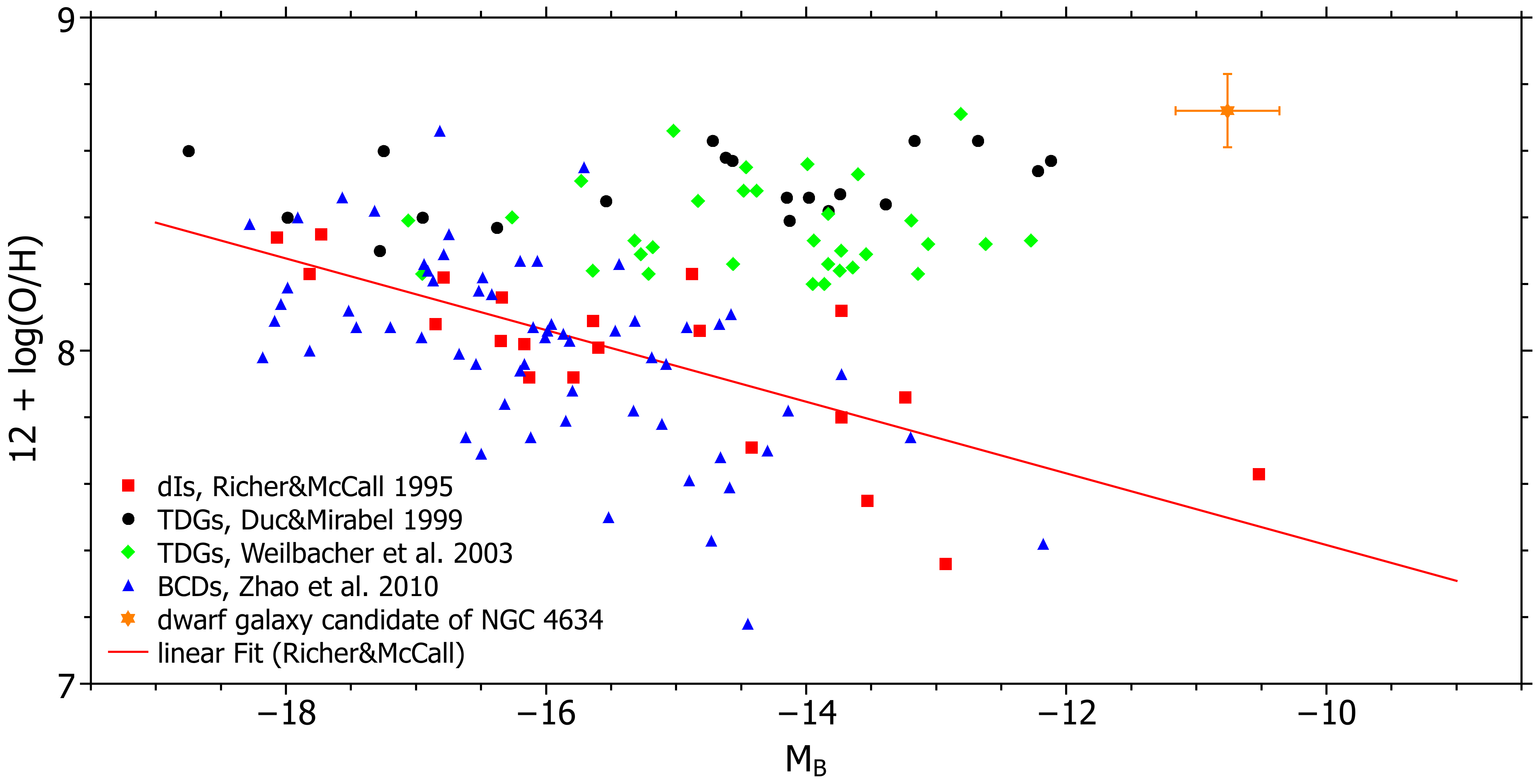}}}
\caption{Luminosity--metallicity relation. The data point of the dwarf galaxy candidate is represented by the orange star. The data of \citet{richermccall1995} of dwarf irregular galaxies (dIs)  are shown as red squares. The fit to that data by \citet{richermccall1995} is shown as the red line. The relation of the data of blue compact dwarfs (BCDs, blue triangles)  is from \cite{zhaoetal2010}. The data of tidal dwarf galaxies (TDG) from \citet{ducmirabel1999} are represented as dark circles, and the data from \citet{weilbacheretal2003} as green diamonds.}
        \label{met-lum-rel-dwarf.png}
\end{center}
\end{figure*}

To compare the calculated L-Z relation for the dwarf galaxy candidate we show in Fig.~\ref{met-lum-rel-dwarf.png} the L-Z relation for five different datasets. 
The data of \citet{richermccall1995} of dwarf irregular galaxies (dIs)  are shown as red squares. The fit to that data by \citet{richermccall1995} is shown as the red line. The relation of the data of blue compact dwarfs (BCDs, blue triangles) from \cite{zhaoetal2010} is similar to the relation of the \cite{richermccall1995} sample. \\
\indent The location of the dwarf galaxy candidate of NGC~4634 is shown with the orange star.
Clearly, the region analyzed here does not follow the relation of the dIs or the BCDs. We  therefore use the data of tidal dwarf galaxies (TDG) from \citet{ducmirabel1999} (dark circles) and the data from \citet{weilbacheretal2003} (green diamonds).\\
\indent Figure~\ref{met-lum-rel-dwarf.png} clearly shows that the abundance of the dwarf galaxy candidate falls within the trend of tidal dwarf galaxies which is mainly due to its high oxygen abundance. As these galaxies are formed from the material of a larger interacting galaxy, they are enhanced in metallicity.  

%________________________________________________________________________________________________________________________________-
\subsection{H{\sc i} data}
The low-resolution H{\sc i} map (Fig.~\ref{HImap}) shows an indication of a spur at the location of the dwarf galaxy candidate. In Fig.~\ref{Velmap} the H{\sc i} velocity map is presented, which shows that the velocity at the location of the dwarf galaxy candidate coincides with the value determined from the spectrum and is also in line with the general rotation of the galaxy.

\begin{figure*}
        \centering
                \includegraphics[width=0.99\textwidth]{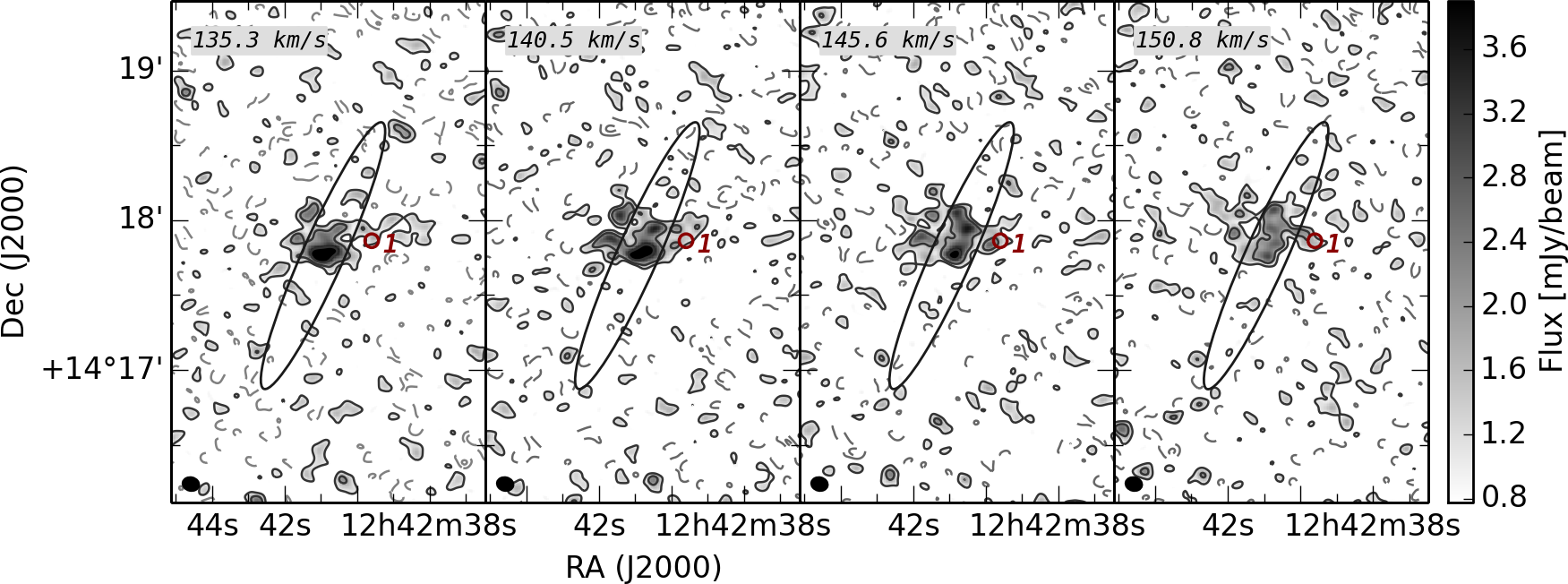}
        \caption{Individual  channel maps of the Hanning smoothed high-resolution H{\sc i} cube. The contours are at -3, -1.5, 1.5, 3, 6, 12 $\times$ 0.62 mJy/beam; negative contours are dashed light gray.}
        \label{HIChan}
\end{figure*}

\begin{figure*}
        \centering
                \includegraphics[width=0.99\textwidth]{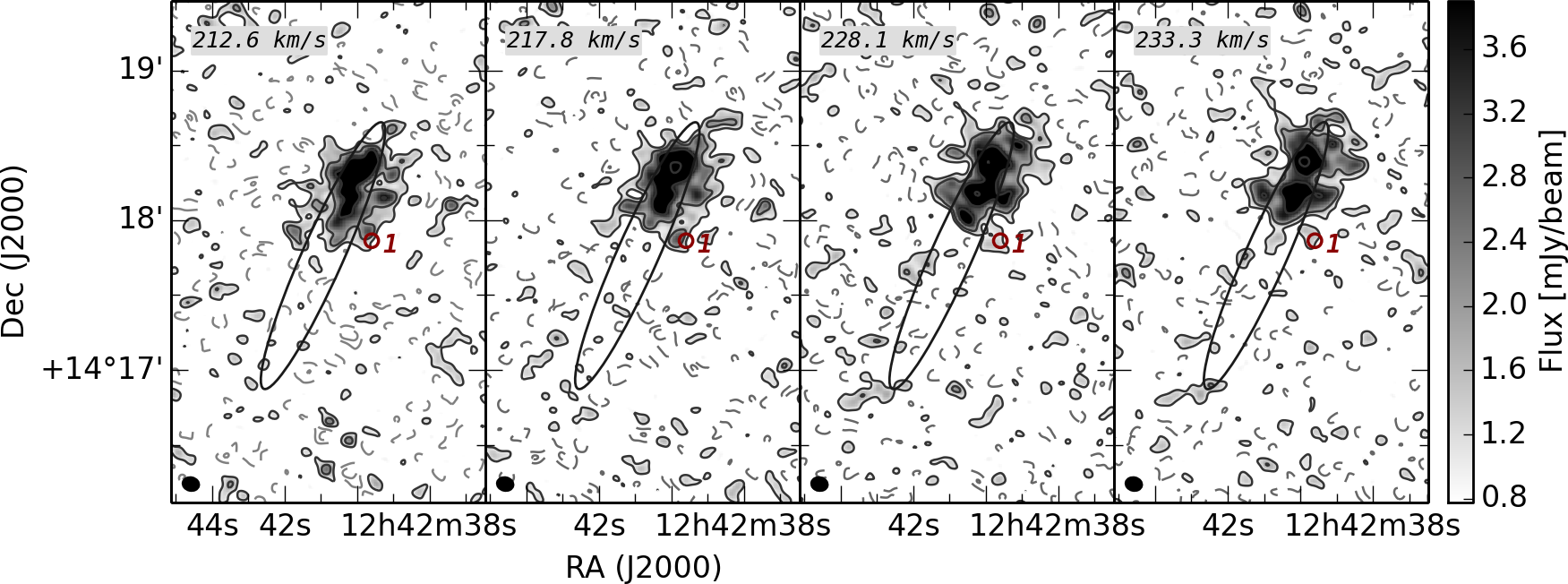}
        \caption{Individual channel maps of the Hanning smoothed high-resolution H{\sc i} cube. The contours are at -3, -1.5, 1.5, 3, 6, 12 $\times$ 0.62 mJy/beam; negative contours are dashed light gray.}
        \label{HIChan200}
\end{figure*}

\begin{figure}
        \centering
                \includegraphics[width=1.\columnwidth]{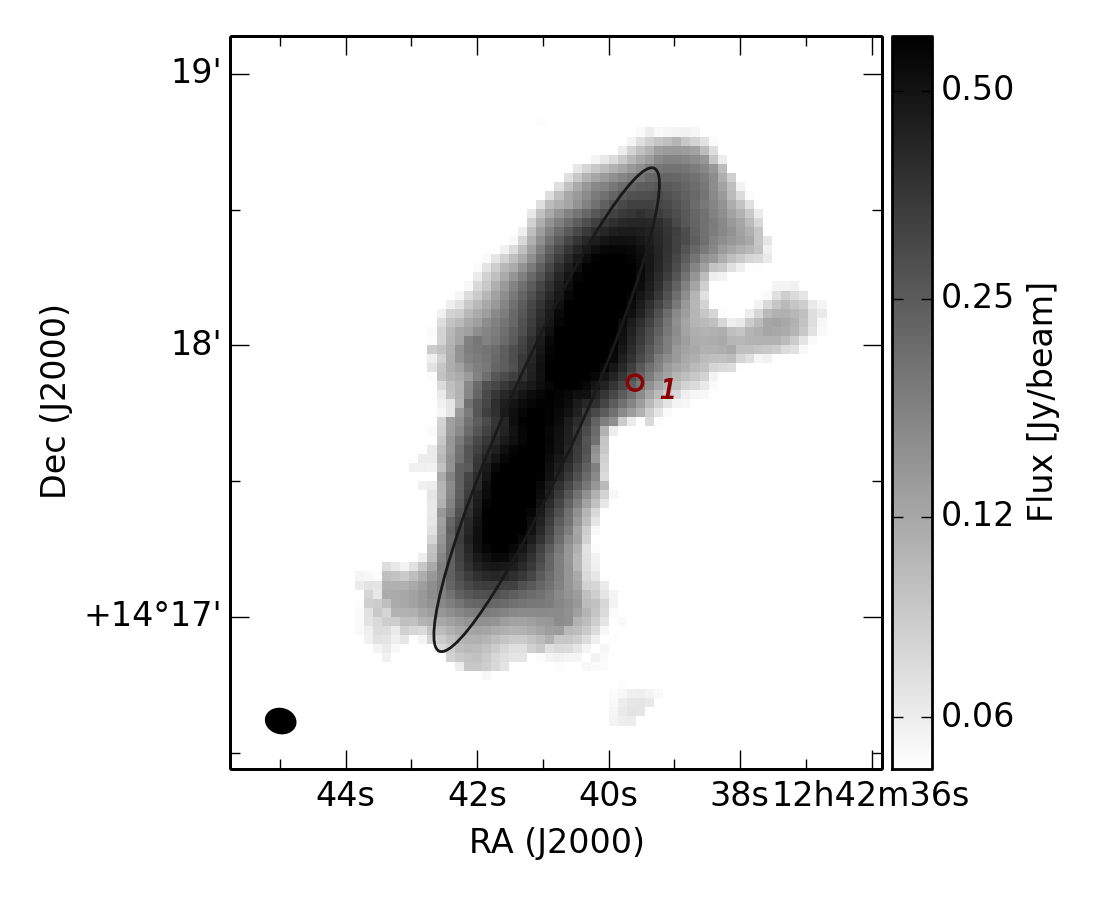}
        \caption{H{\sc i} map in low resolution. The beam is 17\arcsec$\times$18\arcsec\ and is shown in the lower left corner. A 4.3 $\sigma$ mask is applied with $\sigma$~=~11.5 mJy/beam.}
        \label{HImap}
\end{figure}

\begin{figure}
        \centering
                \includegraphics[width=1.\columnwidth]{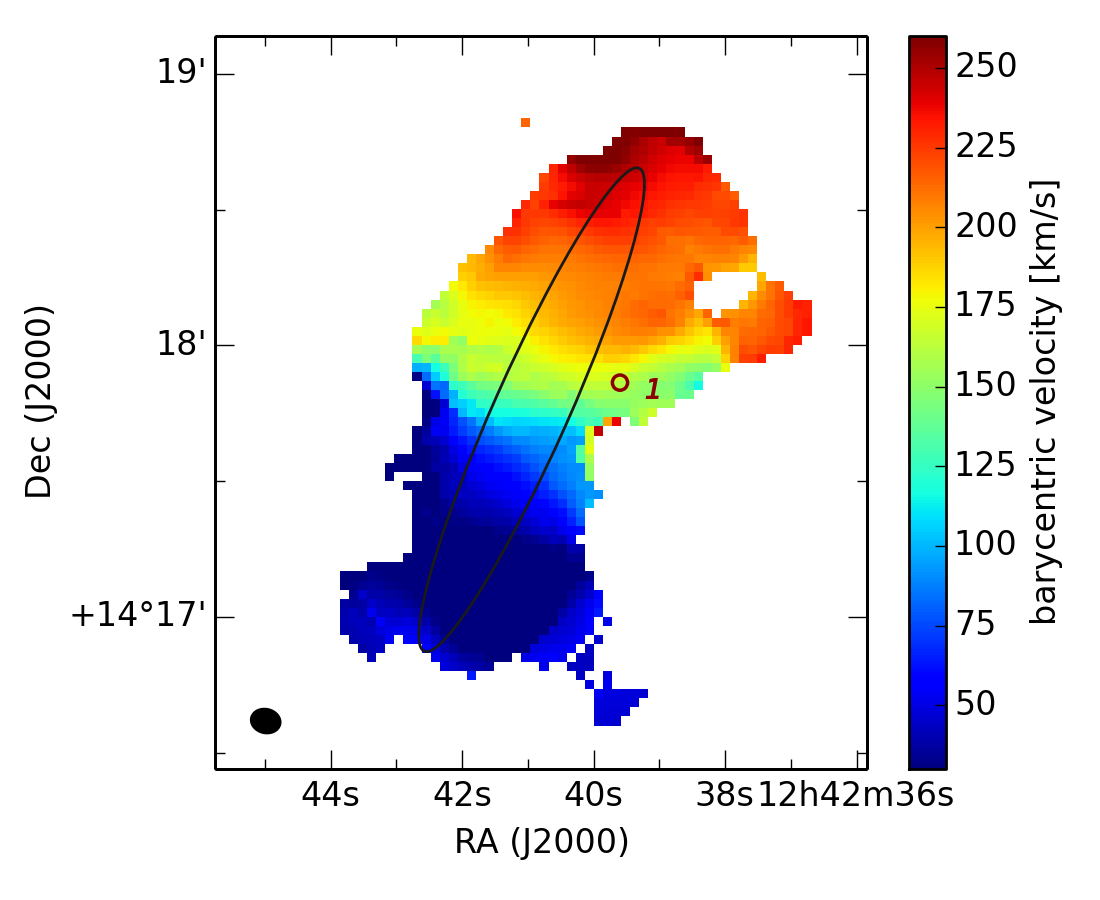}
        \caption{Velocity map in low resolution. The beam is 17\arcsec$\times$18\arcsec\ and is shown in the lower left corner. The same mask as for the H{\sc i} map is used.}
        \label{Velmap}
\end{figure}

We show further channel maps of the high-resolution H{\sc i}-cube around 140 km s$^{-1}$ (Fig.~\ref{HIChan}) and 220 km s$^{-1}$ (Fig.~\ref{HIChan200}). 
From these we see hints of an elongation of the H{\sc i} gas toward the dwarf galaxy candidate. The gas connection to the disk is seen in the two different velocity intervals. In both cases this bridge is seen in four consecutive velocity channels (in Fig. \ref{HIChan200} one channel is omitted). As the Hanning smoothed cube is shown, the velocity resolution is doubled ($\sim$ 10 km s$^{-1}$). Therefore the bridge is visible in two independent channels. In Fig.~\ref{HIChan} the H{\sc i} gas shows elongation perpendicular to the major axis of the disk of NGC~4634 clearly showing a disturbance in the H{\sc i} distribution. The dwarf galaxy candidate is located in a dense spot of the H{\sc i} gas distribution but with a low level of significance.

\subsection{Total mass of NGC~4634}
We further determined the HI mass and dynamical mass of NGC~4634. We fit a simple tilted ring model to the galaxy using the Fully Automated TiRiFiC \citep[FAT,][]{kamphuis2015soft}. Visual inspection of the output model shows that the fit is reasonable. The heliocentric systemic velocity of NGC~4634 is found to be 115~km~s$^{-1}$. This deviates significantly from the value reported by \citet{rotv46341993}. However, a comparison between their Table~2 and Fig.~4 shows immediately that their reported velocity values for NGC~4633 and NGC~4634 have been swapped in their Table~2. Our value compares well to the systemic velocity reported for NGC~4633. The observed total flux in our GMRT observation is 3.9~Jy~$\times$~km~s$^{-1}$, which is significantly less than values reported in the literature; for example, the Arecibo Legacy Fast ALFA Survey \citep{haynesetal2011} finds a total flux of 7.6~Jy~$\times$~km~s$^{-1}$. This indicates that our observations are not sensitive enough to detect low column density gas. This means that at a distance of 19.1~Mpc the total mass of the galaxy is in the range of 3.3--6.5~$\times$~10$^{8}$~M$_{\odot}$ of H{\sc i}.\\
\indent The FAT model shows a symmetric rotation curve which flattens off at $\sim$~135~$\pm$~9.8~km~s$^{-1}$. At the  radius corresponding to a density of 1~M$_{\odot}$~pc$^{-2}$ of  6.2~kpc (D$_{\rm HI}$~=~134") this translates to a total dynamical mass of 1.3~$\times 10^{10}$~M$_{\odot}$.

%%%%%%%%%%%%%%%%%%%%%%%%%%%%%%%%%%%%%%%%%%%%%%%%%%%%%%%%%%%%%%%%
\section{Discussion}
All important properties of the dwarf galaxy candidate based on the analysis done in this work are presented and discussed in this section.

\subsection{Dwarf galaxy or star-forming patch}
\label{4.1}
To classify faint star-forming objects with no information about the gas kinematics as a galaxy is not straightforward. Even some of the faint dwarf galaxies around the Milky Way cannot be confirmed as galaxies by certain definitions of galaxies. Here we define a galaxy as a gravitationally bound collection of stars whose properties cannot be explained by a combination of baryons and Newton's laws of gravity \citep[see][]{willmanstrader2012}. This definition includes a dark matter halo. This is not fulfilled automatically by TDGs, especially as there are observational hints which are supported by simulations that TDGs do not contain dark matter or are deficient in dark matter \citep{Duc2012,lellietal2015}. Therefore, we summarize the main characteristics of the   dwarf galaxy candidate analyzed here, and if possible compare them with the known characteristics of dwarf galaxies, tidal dwarf galaxies, and globular clusters to get hints on the classification.

In the spectra we see galaxy continuum at the location of the dwarf galaxy candidate's H{\sc ii} region. Its core size in H$\alpha$ is about 90~pc in diameter, including the diffuse emission, and the star's size is 550~pc in diameter. This is quite small for a dwarf galaxy, but too big for a usual globular cluster of a few pc in the Milky Way \citep{harris1996}. Nevertheless, it is comparable to several dwarf galaxies in the Local Group \citep{mateo1998}.

The projected distance of the dwarf galaxy candidate to its host galaxy NGC~4634 is 1.4 kpc. This distance is quite small. The projected distance  merely provides a lower limit, and the true distance between NGC~4634 and the dwarf galaxy candidate could be higher. However, as they have similar velocities we  expect them to have similar distances, and thus the true separation between the dwarf galaxy candidate and NGC~4634 is not expected to be orders of magnitude larger than the projected distance.
 
The photometric analysis shows that the galaxy appears very young. As it is seen in the HST F625W and H$\alpha$ images, the dwarf galaxy candidate is actively producing stars and contains already-formed stars. Comparing the computed Lyman-$\alpha$ photons to \citet{panagia73}, a few O4 stars or several hundred  B0 stars are able to produce the required number of photons. The number of log(L(H$\alpha$))~=~38.88~$\pm$~0.16~erg~s$^{-1}$ is, considering the low absolute $B$-band magnitude of M$_B$~=~$-$10.76~mag, high compared to local dwarf and tidal dwarf galaxies analyzed by \citet{croxall2009}, where most dwarf galaxies comparable in the absolute flux show significantly lower H$\alpha$ fluxes of log(L(H$\alpha$))~$<$~36.6~erg~s$^{-1}$. The derived SFR from UV (1~$\pm$~0.1)~$\times$~10$^{-3}$~M$_\odot$~yr$^{-1}$ is relatively high compared, for example,  to the dwarf irregular galaxy DDO210 of the local group with M$_B$~=~$-$10.76~mag and the UV SFR~=~1.58 $\times$ 10$^{-4}$ M$_\odot$ yr$^{-1}$ from \citet{leeetal2009}. The computed stellar mass of $1.35 \times 10^6 M_\odot$ is low in comparison to other dwarf galaxies and tidal dwarfs \citep[e.g.,][]{kavirajetal2012}, but high for globular clusters \citep[e.g.,][]{vanzellaetal2017} placing the stellar mass of the dwarf galaxy candidates in the intermediate regime between dwarf galaxies and globular clusters.

The absolute $B$-band magnitude (-10.76 mag) is  low in comparison to the presented samples of other dwarf galaxies, which is probably a selection effect as faint objects are not detected easily. Nevertheless, in the \citet{richermccall1995} sample, one dwarf galaxy (GR 8) exists with a comparable M$_B$. Furthermore, it is comparable to faint Milky Way and local group dwarf galaxies, as seen above.

With all this, the conclusion of \citet{rossaetal2008} that the dwarf galaxy candidate is likely a galaxy might be true. Nevertheless, like for other faint dwarf galaxies, the uncertainties of a galaxy classification are large \citep{willmanstrader2012} and the close distance to NGC~4634 is still peculiar. With the data presented here we cannot establish that it is gravitationally bound, let alone whether it is embedded in its own dark matter halo, and hence drawing a strong conclusion on the dwarf galaxy candidate being a galaxy must be postponed until such data can be obtained.

\subsection{Observational hints to the origin of the dwarf galaxy candidate}
Because of the similar velocities of the dwarf galaxy candidate and the host NGC~4634 (from long-slit spectra and H{\sc i}), they seem to be coupled to each other. The mean oxygen abundance of the dwarf galaxy candidate of 8.72 is similar  to that of the host galaxy NGC~4634. Comparing the L-Z relation of isolated dwarf galaxies with the dwarf galaxy candidate of NGC~4634, it is obvious that they are not correlated. The dwarf galaxy candidate is located in the upper part due to a higher metallicity and seems to follow the correlation of tidal dwarf galaxies. The elongation in the appearance of the dwarf galaxy candidate is perpendicular to the major axis of NGC~4634 in the DSS-image. This seems to be confirmed by the low-resolution H{\sc i} map where a tentative spur in the same direction is visible. The high-resolution channel maps show a dense spot at the location of the dwarf galaxy candidate. This leads to the conclusion that, whether it is  a galaxy or not, the material from which the dwarf galaxy candidate was created came from the disk of NGC~4634 or is at least equally chemically enriched.

\subsection{Possible origins of the dwarf galaxy candidate}
Origin 1: Material from NGC~4634\\
The oxygen abundance of the extraplanar region is high in comparison to other dwarf galaxies and comparable to the host galaxy NGC~4634. The position of the dwarf galaxy in the L-Z relation indicates that its material comes  from NGC~4634. The analyzed dwarf galaxy candidate is located very close to NGC~4634 and seems to be young. This could be explained by two different scenarios pulling or pushing the material out of the midplane of NGC~4634 into the halo. This might be outflow driven or tidal.

\textit{Outflow driven:}  The region could be a star-forming region in the lower halo and no dwarf galaxy at all. The material originating from NGC~4634 was expelled from the galaxy disk via outflow processes. The main process of outflow models is thermal pressure from supernova-driven energy, where SNe explosions shock heat the interstellar medium (ISM) \citep[e.g.,][]{larson1974}. \citet{nath2013} suggested that correlated supernovae from OB associations in molecular clouds in the central region can drive powerful outflows. With the other components of the ISM also playing an important role, e.g., \citet{samuietal2018}  conclude that CR-driven outflows can go to a larger radii compared to a purely thermally driven outflow with same total energy input.
With the outflow scenario the expelled gas condenses in the halo and forms new stars. A similar scenario in an AGN driven outflow was recently discussed by \citet{Maiolinoetal2017} for another galaxy. This scenario cannot be ruled out with the current data.

\textit{Tidal:} The findings could be explained by a tidal dwarf scenario. \cite{zwicky56} argued long ago that tidal forces can tear long tails of stars and gas from the bodies of interacting disk galaxies, and that this debris may include self-gravitating objects, which could become small galaxies. Numerical simulations of encounters between disk galaxies show that dwarf systems can form in material that is drawn out during an encounter \cite[e.g.,][]{barnesherquist92}. 

The key element is that the building material of TDGs used to belong to a larger parent galaxy, and thus it is pre-enriched \citep{Duc2012}. Therefore, dwarf means that the object born in tidal features should have the size and mass of a dwarf galaxy. In observations that would imply that TDGs have inherited the metal content of interstellar medium from their parents. Their metallicity tells about the past chemical enrichment of their parents, and is therefore not correlated with their actual mass, contrary to conventional galaxies. Additionally, their dust content and molecular gas content, as traced by CO, is higher than in regular star-forming dwarf galaxies \citep{Duc2012}.

The high oxygen abundance and the other characteristics of the dwarf galaxy candidate could be explained by the tidal dwarf scenario. It is therefore possible that the dwarf galaxy was formed in the spur as a result of  tidal interactions with the neighboring galaxy, namely NGC~4633. However, the low-resolution H{\sc i} image of the NGC~4634 and NGC~4633 does not show any cold gas between the two galaxies. This might be due to sensitivity limits (1$\sigma~\sim~3~\times~10^{19}$~cm$^{2}$; see also Section~3.7);  a comparison between single-dish data (Haynes et al. 2011) and our GMRT observations  indicates we are missing a significant amount of emission. However, both disks are clearly detected and show little to no signs of being disturbed. On the other hand, the F625W image (Fig. 7) shows clear signs of the stellar disk being disturbed, indicating past interactions. Hence,  NGC~4634 has clearly undergone recent interaction even though it remains unclear whether this was caused by NGC~4633.

Without more detailed kinematical information on the dwarf galaxy candidate it is impossible to distinguish between the two possible forces, outflow driven and tidal.\\

\noindent Origin 2: Dwarf galaxy\\
In hierarchical structure formation scenarios such as $\Lambda$ cold dark matter ($\Lambda$CDM), it is expected that dwarf galaxies  merge into big galaxies. Especially in the group environment of the Virgo cluster, this could be the case. Thus, the extraplanar region could also be a dwarf galaxy of the Virgo Cluster that is gravitationally coupled to NGC~4634, but not necessarily a tidal dwarf. The analysis performed here shows that, except for its metallicity, the structural parameters of the dwarf galaxy candidate are consistent with it being a dwarf galaxy (see discussion in Section~4.1
and, e.g., \citealt[]{misgeldetal2011} for further comparisons). However, it is rather puzzling as to why the candidate should then have a metallicity comparable to NGC~4636. Could it be due to the cluster environment that both galaxies reside in, assuming there is more interaction between galaxies and intermixing of the gas happening in a cluster environment? The argument against this comes from \citet{leeetal2003}, who concluded that at a given optical luminosity there is no systematic difference in oxygen abundance between a sample of Virgo star-forming dwarfs and a sample of nearby star-forming dwarfs. Therefore, this scenario is unlikely.\\

\noindent Origin 3: Gas cloud\\
The analyzed dwarf galaxy candidate could have been a cloud in the Virgo cluster.
The material could originate from the outer part of galaxies torn by tidal forces in
an interaction or ram pressure stripping as suggested for the Virgo clouds in \citet{Kentetal2009}. It furthermore could originate from gas clouds with very low star formation. This  free floating gas would likely be metal poor, as the classical example of the Virgo HI cloud shows \citep{salzer1991}. We expect then a relatively low oxygen abundance, which is not the case. In contrast to this, with the first origin of a cloud, the material would have comparable oxygen abundances to other galaxies in the cluster. We therefore cannot rule out for sure that the dwarf galaxy candidate originates from another galaxy of the Virgo Cluster. Nevertheless, the velocity map suggests that the velocity of the dwarf galaxy candidate is not disturbed and in line with the general rotation of NGC~4634, which would  probably not be the case for an infalling gas cloud.

 %________________________________________________________________________________________________________________________
\section{Summary and conclusion}
In this paper we present multifrequency data of a dwarf galaxy candidate and its edge-on host NGC~4634. With optical long-slit spectra we calculate the oxygen abundances of one disk H{\sc ii} region and of a dwarf galaxy candidate's H{\sc ii} region. The comparable oxygen abundances hint at the material of the dwarf galaxy candidate being pre-enriched and that it originates from the disk of NGC~4634. This is confirmed by its location in a L-Z diagram, which shows that its metallicity is much higher than expected for a dwarf galaxy that evolved in relative isolation. The images of r-band, $B$-band, and the H$\alpha$-luminosity show that the dwarf galaxy candidate is star forming and approximately 550\,pc in diameter. The heliocentric velocities of the dwarf galaxy and its host NGC~4634 are similar and indicate that the galaxies are connected. The H{\sc i} data give the hint that a spur of NGC~4634 is located at the same position as the dwarf galaxy candidate. We furthermore determined the total dynamical mass of NGC~4634 to be 1.3~$\times 10^{10}$~M$_{\odot}$ from the   GMRT data analyzed here.\\  
\indent With the above discussion one possible origin is the tidal dwarf scenario where the dwarf galaxy candidate formed in the spur of NGC~4634 from disk material. However, the data presented in this paper cannot rule out other forces that separated the gas in the star-forming region from the disk of NGC~4634. In any case, it is clear that the halo of NGC~4634 contains a star-forming object of significant size that is formed from gas stripped from NGC~4634, which is an excellent laboratory for further studies in terms of galaxy evolution.

\begin{acknowledgements}
We thank the anonymous referee for the helpful and constructive comments. Furthermore, we thank Pierre-Alain Duc for interesting discussions about this work. We thank Anika Beer for helping perform the pPXF.
 
This research was suported in part by the DFG (German Research Foundation) research unit FOR1048. This research has made use of the ``Aladin sky atlas'' developed at CDS, Strasbourg Observatory, France, and has made use of IRAF, the NASA's Astrophysics Data System Bibliographic Services, the GOLDMine Database, and the NASA/IPAC Extragalactic Database (NED) which is operated by the Jet Propulsion Laboratory, California Institute of Technology, under contract with the National Aeronautics and Space Administration. This research has used data provided by the GALEX mission, the SDSS, and UKIDSS. GALEX is a NASA small explorer, launched in April 2003. It is operated for NASA by Caltech under NASA contract NAS5-98034. Funding for the Sloan Digital Sky Survey IV has been provided by the Alfred P. Sloan Foundation, the U.S. Department of Energy Office of Science, and the Participating Institutions. SDSS-IV acknowledges support and resources from the Center for High-Performance Computing at the University of Utah. The SDSS web site is www.sdss.org. SDSS-IV is managed by the Astrophysical Research Consortium for the Participating Institutions of the SDSS Collaboration including the 
Brazilian Participation Group, the Carnegie Institution for Science, Carnegie Mellon University, the Chilean Participation Group, the French Participation Group, Harvard-Smithsonian Center for Astrophysics, Instituto de Astrof\'isica de Canarias, The Johns Hopkins University, Kavli Institute for the Physics and Mathematics of the Universe (IPMU) / 
University of Tokyo, the Korean Participation Group, Lawrence Berkeley National Laboratory, Leibniz Institut f\"ur Astrophysik Potsdam (AIP),  
Max-Planck-Institut f\"ur Astronomie (MPIA Heidelberg), Max-Planck-Institut f\"ur Astrophysik (MPA Garching), 
Max-Planck-Institut f\"ur Extraterrestrische Physik (MPE), 
National Astronomical Observatories of China, New Mexico State University, 
New York University, University of Notre Dame, 
Observat\'ario Nacional / MCTI, The Ohio State University, 
Pennsylvania State University, Shanghai Astronomical Observatory, 
United Kingdom Participation Group,
Universidad Nacional Aut\'onoma de M\'exico, University of Arizona, 
University of Colorado Boulder, University of Oxford, University of Portsmouth, 
University of Utah, University of Virginia, University of Washington, University of Wisconsin, 
Vanderbilt University, and Yale University. The UKIDSS project is defined in Lawrence et al. (2007). UKIDSS uses the UKIRT Wide Field Camera (WFCAM; Casali et al. 2007). The photometric system is described in Hewett et al. (2006), and the calibration is described in  Hodgkin et al. (2009). The pipeline processing and science archive are described in Irwin et al. (2009, in prep.) and Hambly et al. (2008).
\end{acknowledgements}

\bibliography{BibliographyN4634}
\bibliographystyle{aa}

\end{document}